\documentclass[aps,pre,onecolumn,superscriptaddress,floatfix]{revtex4}%

\usepackage{times}
\usepackage{amssymb}
\usepackage{amsbsy}
\usepackage{amsmath}
\usepackage{amsbsy}
\usepackage{latexsym}

\usepackage[dvipsone]{graphicx}

\usepackage{subfigure}

\newcommand{\br}{\mathbf{r}}
\newcommand{\bp}{\mathbf{p}}
\newcommand{\bQ}{\mathbf{Q}}
\newcommand{\bu}{\mathbf{u}}
\newcommand{\bv}{\mathbf{v}}
\newcommand{\bx}{\mathbf{x}}
\newcommand{\x}{\bx}                         % set of coarse-grained variables
\newcommand{\z}{\mathbf{z}}                  % point in micro phase space
\newcommand{\balpha}{\boldsymbol{\alpha}}
\newcommand{\bkappa}{\boldsymbol{\kappa}}    % velocity gradient
\newcommand{\bsigma}{\boldsymbol{\sigma}}    % stress tensor
\newcommand{\bPi}{\boldsymbol{\Pi}}
\newcommand{\bLambda}{\boldsymbol{\Lambda}}  % Lagrange multiplier
\newcommand{\blambda}{\boldsymbol{\lambda}}
\newcommand{\kb}{k_{\rm B}}                  % Boltzmann's constant
\newcommand{\mom}{g}                         % momentum density
\newcommand{\bmom}{\mathbf{\mom}}
\newcommand{\macro}{c}                       % macro variable: conformation tensor
\newcommand{\bmacro}{\mathbf{\macro}}
\newcommand{\bcvlasis}{\bmacro}              % before: {\tilde{\bmacro}}
\newcommand{\cvlasis}{\macro}                % before{\tilde{\macro}}
\newcommand{\rhox}{\rho_{\x}}                % generalized canonical distribution
\newcommand{\Umf}{U_{\rm mf}}                % potential of mean force
\newcommand{\en}{\epsilon}                   % energy density
\newcommand{\Ree}{\mathbf{R}_{\rm ee}}       % end-to-end vector
\newcommand{\Nch}{N_{\rm ch}}                % number of chains
\newcommand{\Vol}{\mathcal{V}}               % volume
\newcommand{\taus}{\tau_{\rm s}}             % separating time scale
\newcommand{\taup}{\tau_{\rm p}}             % longest relaxation time (in Wi)
\newcommand{\Wi}{\mathrm{Wi}}                % Weissenberg number
\newcommand{\ncg}{n_{\rm CG}}                % number of coarse-grained variables
\newcommand{\nsample}{n_{\rm s}}             % number of samples 

\newcommand{\Deltalambda}{\delta\blambda}
\newcommand{\bbf}[1]{\mbox{\boldmath{$#1$}}}
\newcommand{\ave}[1]{\langle #1 \rangle}     % average
\newcommand{\avex}[1]{\langle #1 \rangle_x}
\begin{document}

\title{Multiscale Modeling and Coarse Graining of Polymer Dynamics:\\
Simulations Guided by Statistical Beyond-Equilibrium Thermodynamics
\footnote{to appear in P.D.~Gujrati and A.L.~Leonov (Ed.), \textit{Modeling and Simulations in Polymers}, Wiley 2009, in press.}}

\newcommand{\ETH}{Polymer Physics, ETH Z\"urich, Department of Materials, CH-8093 Z\"urich, Switzerland}
\newcommand{\UPA}{Department of Chemical Engineering, University of Patras and FORTH-ICE/HT, Patras, GR 26504, Greece}

\author{Patrick Ilg}
\affiliation{\ETH}
\author{Vlasis Mavrantzas}
\affiliation{\UPA}
\author{Hans Christian \"Ottinger}
\affiliation{\ETH}
\date{\today}

\maketitle

%%%%%%%%%%%%%%% Introduction %%%%%%%%%%%%%%%%%%%%%%%%%%%%%%%%%%%%%%%%%

\section[Motivation and goals]{Polymer dynamics and flow properties we want to understand: Motivation and Goals}

\subsection{Challenges in polymer dynamics under flow}
Polymer molecules differ from simple fluids in several aspects: they are extremely diverse in structure 
(they can have a linear, branched, ring-like, or block copolymer structure), 
they can be characterized by a molecular weight distribution, and they are capable of exhibiting a huge number 
of configurations implying that a large number of degrees of freedom should be accounted for in any molecular 
modeling approach. As a result, polymers exhibit properties which are totally distinct from those of the simpler 
Newtonian liquids. The drag reduction phenomenon (the substantial reduction in pressure drop during the turbulent 
flow of a Newtonian liquid when a very small amount of a flexible polymer is added), their unique rheological 
properties (shear thinning and normal stress differences in simple shear, strain hardening in elongation, 
complex viscosity, anisotropy in thermal conductivity and diffusivity), and a plethora of other phenomena 
associated with their elastic character are only a few manifestations of the departure of their behavior from 
the Newtonian one \cite{DealyLarson,DPL12}. 
Of particular importance from a mechanical or fluid dynamics point of 
view is their viscoelasticity quantifying the irreversible conversion of the work needed for their deformation 
to heat loss but also their capability to store part of this work as elastic energy. It is a property closely 
related to the multiplicity of time and length scales characterizing dynamics and structure in these fluids. 
Thus, even in the viscous regime ($\Wi\ll 1$, where $\Wi$ is the Weissenberg number empirically defined as 
$\Wi=\taup\dot{\gamma}$ with $\taup$ being the longest relaxation time and $\dot{\gamma}$ the flow rate),
the flow can still be strong enough for several degrees of freedom not to be close to equilibrium giving 
rise to interesting rheological properties also there \cite{DPL12}, especially for 
high molecular weight polymers. 

Understanding relaxation processes and structure development occurring over these multiple scales is a 
prerequisite for deriving reliable constitutive equations connecting 
the stresses developing in these materials in terms of the imposed flow kinematics and certain molecular 
parameters or functions, and for computing polymer flows 
\cite{Owens_computrheo,Keunings_advances,Malikin_polyrheo,McLeish_science,johannes,schoonen2,verbeeten,hassel}. 
It is only through a comprehensive understanding 
across scales that one can hope to build the relationship between polymer molecular structure, conformation and 
architecture and macroscopic rheological response. In addition to experiments and theories, molecular 
simulations can play a significant role by providing high resolution calculations especially on the crossover from 
small-to-intermediate scales. This chapter is devoted to a brief discussion of some of the emerging multi-scale 
simulation approaches in the recent research literature on nonequilibrium systems, 
with emphasis on those based on well-founded theoretical frameworks. 
Our goal is to demonstrate that, with the help of and guidance from recent 
advances in the field of nonequilibrium statistical mechanics and thermodynamics, this highly demanding endeavor 
(modeling across scales) can lead to simulation methodologies that have been elevated from simple, brute-force 
computational experiments to systematic tools for extracting complete, redundancy-free, and consistent 
coarse-grained information for the flow dynamics of polymeric systems\cite{hco_lessons}.

\subsection{Modeling polymer dynamics beyond equilibrium} \label{sec:polymodels}

\begin{figure}
 \includegraphics[width=8cm]{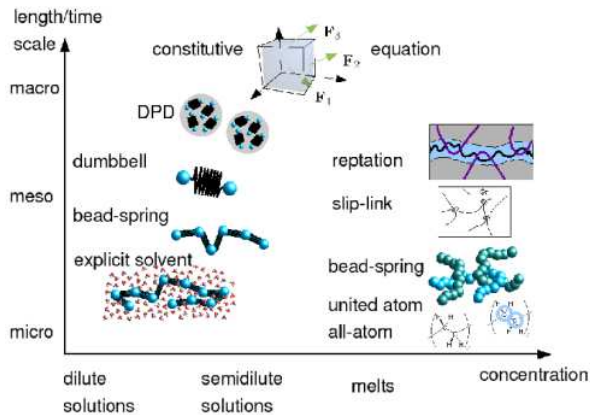}
 \caption{Different models of polymer dynamics are shown schematically for solutions and melts.}
 \label{models_schematic.fig}
\end{figure}

Describing macromolecular configurations under nonequilibrium conditions is an extremely difficult problem, 
which usually requires simplified models for analytical or numerical studies \cite{DPL12}; 
such simple models have contributed enormously to our understanding of polymer rheology and 
mechanics. For a review on proposed models, see
e.g.~\cite{mkbook,Kremer_ModelReview,doibook,birdwiest_review,birdhco_review}, 
while for a review on available simulation tools addressing different time and length scales, see 
\cite{Colbourn,Binder_MCMDpoly,GlotzerPaul,Kremer_entangledsimu,Kremer_review03,Doros_simureview,plathe_CGinLNP}. 
Fig.~\ref{models_schematic.fig} shows a schematic of the pertinent models for polymer solutions and melts 
depending on the length- and time-scale of interest.

In general, examples of macroscopic constitutive equations employed to calculate polymer flow behavior include 
e.g.~conformation tensor models like the Giesekus, Maxwell and FENE models but also the more recently
proposed Pom-Pom, CCR and Rolie-Poly models 
\cite{birdwiest_review,birdhco_review,pompom,wapperom_pompom,likhtman_rolie,marrucci_constitutive01}. 
For an overview, see \cite{Owens_computrheo,Keunings_advances} as well as the contributions by A.N.~Beris and 
E.~Mitsoulis in this volume. Most of these constitutive equations have been derived (or inspired) by simple 
mechanical models of polymer motion. They are mesoscale, kinetic theory models (based, e.g., on the dumbbell, 
FENE, bead-spring chain, and bead-rod chain analogues) capable of accounting for some important aspects of 
polymer dynamics, like chain stretching, non-affine deformation, diffusion, and hydrodynamic drag forces, 
either separately or all together \cite{pavlos_JOR1}. In polymer solutions, in particular, one should account 
for the solvent-mediated effect between beads (known as hydrodynamic interaction), which is usually modeled with 
the hydrodynamic resistance matrix \cite{DPL12,kimkarrila}. 
Recently, an efficient simulation of solvent dynamics with 
the help of either the lattice Boltzmann 
\cite{saurobook,duenweg_LBreview} or the stochastic rotation dynamics
\cite{SRD_colloidinshear,SRD_review} methods has been proposed with a
suitable coupling of the bead dynamics. Besides their efficiency, these combined methods also incorporate 
non-uniformities and fluctuations in internal flow fields.

Usually, the mesoscopic, kinetic models are considered to be well-suited for predicting dynamical properties of 
polymer solutions on macroscopic scales. Details of the fast solvent dynamics are in
most cases irrelevant for macroscopic properties. Exceptions are
polyelectrolytes, where the motion of counterions in the solvent can
have a major influence on polymer conformation. Therefore, more
microscopic models of polyelectrolytes with explicit counterions are
sometimes employed \cite{holm_PEreview} (see also the contribution by M.~Muthukumar in this volume). 
Another exception is the dynamics of individual biopolymers, e.g.~protein folding, which is modeled with an all 
atomistic model including an explicit treatment of the (water) solvent molecules \cite{Shea_proteinfold}.

For typical polymer melts, dumbbell models are inappropriate since
they fail to account for the essential role of entanglements on the long-time dynamics. 
Successful mesoscopic, mean-field descriptions of entangled
melts are offered by the reptation models \cite{doibook}. Several modifications of
the original Doi-Edwards-de Gennes reptation theory have been proposed in the last years in order to improve 
comparison to experiments in the linear and nonlinear flow regimes 
\cite{graham_tubeimproved,likhtman_testtube,marrucci_multimode,hco117,hco119,hco123,hco129,meadlarsondoi}. 
Recently, slip-link models have also been proposed  
\cite{jay_sliplink,larson_sliplink,jay_sliplink2,likhtman_sliplink,doi_sliplink,marrucci_sliplink,marrucci_sliplink2}, 
providing a slightly more detailed description of entanglements, and which agree well with
available experimental results.

At the microscopic level, polymer melts are modeled as multi-chain systems, 
see e.g.~\cite{DPL12,mkbook,Rigby_polymodels,Kremer_ModelReview}. 
For example, all-atom or united atom force fields, accounting explicitly for bond angle bending and 
torsion angle contributions (in addition to bond stretching and intermolecular interactions) 
\cite{Siepman,Toxvaerd}, are available. 
Different united-atom force fields are reviewed and compared e.g.~in 
\cite{Rigby_polymodels,Rousseau,MatticeSuter}. 
From such detailed atomistic molecular dynamics (MD) simulations, 
the linear viscoelastic properties can be computed by 
Green-Kubo relationships \cite{kubobook2,Barrat_Rouse,Briels_CGreview}. 
Also less detailed bead-spring models are available; 
a prototypical model is that of Kremer and Grest \cite{KremerGrest} (or variants thereof \cite{loose}), 
which neglects chemical details and instead focuses on the interplay between chain connectivity and
excluded volume effects.
The non-linear regime can be studied by nonequilibrium molecular dynamics (NEMD) simulations, 
that directly address flow effect on polymer structure and conformation,
both in shear and planar elongation 
(see e.g.~\cite{mkbook,loose,todd_nemdreview,toddreview,mk_crossover,brian_NEMD_1,brian_NEMD_2,chunggi_NEMD_1,chunggi_NEMD_2} 
and references therein). 
They are based on flow-adapted boundary conditions such as those proposed by Lees-Edwards for planar shear 
\cite{Lees_Edwards} and by Kraynik-Reinelt for planar elongation \cite{Kraynik_Reinelt}.

\subsection{Challenges in standard simulations of polymers in flow} \label{sec:challenges}

Despite the enormous advances in the field of molecular simulations 
\cite{Keunings_advances,Binder_MCMDpoly,Kremer_review03,Doros_simureview,Doi_challenge,Richter_dynsfac,Cummings_NEMDandExp,Tuzun_advancespolysimu}, 
predicting the macroscopic flow properties of polymers from their underlying microstructure presents 
still major challenges \cite{Ober_polychallenge}. 
Available MD and NEMD algorithms can address only time scales on the order of a few microseconds at most, 
implying that only short up to moderately entangled polymers can be studied in full atomistic detail in a 
brute-force manner. Extending the simulations to longer, truly entangled polymers is a first big task. 
Extending these flows to mixed or inhomogeneous flows is another big challenge. 
Among others, such a development would help understand the 
origin of interfacial slip and its mode (localized slip versus global slip) in the flow past a solid substrate. 
On the other hand, with the introduction of the revolutionary set of chain connectivity altering moves, extremely 
powerful Monte Carlo algorithms have been developed which have helped overcome the issue of the thermal 
equilibration of long polymers even at beyond equilibrium conditions 
\cite{vlasis_EB,vlasis_Encyclopedia,nikosPRL2002,Kremer_equilibrate,Doros_CGreview}. 
With the help of the end-bridging and double bridging moves, for example, truly long polyethylene 
(linear and branched) and polybutadiene systems have been equilibrated over the last years, 
which opened also the way to their topological analysis for the identification of entanglements 
\cite{martin_primitivepath,Everaers_primitivepath,Everaers_science,Siheung,hco142,hco145,brian_jnnfm}.

Arguably the biggest challenge in polymer simulation under nonequilibrium conditions is to build 
well-founded multi-scale tools which can bridge the gap between microscopic information and 
macroscopically manifested viscoelastic properties, preferably through a constitutive equation 
founded on the microscopic model \cite{hco_lessons,Doi_challenge}.
Simulations of metals face the same problem where again the objective is concurrent length scale 
simulations 
\cite{Tadmor_MRS,Kaxiras_multiscalemetals}; to some extent, it is also relevant to simple fluids 
\cite{coveney_prl}. 
For polymers, additional motivation stems from the increased interest in 
polymer mixtures and interfaces \cite{Milner_polyinterface} resulting in morphology development at 
the nanoscale.

Here, we aim to shortly review some recent coarse-graining and multi-scale methods 
(see 
\cite{guenza_reviewCG,Doros_CGreview,Kremer_CGsimu08,Kremer_reviewMultiscale,Briels_CGreview,Kremer_multisimu,plathe_CGinLNP,Baschnagel_CGreview}
and also \cite{Nielaba_bridgingbook,eijnden_progress,Voth_CGbook,Murtolo_MultiScaleSoftMatterreview} for recent reviews of 
such methods for polymers and in a more general context, respectively), 
but also to put forward some new ideas for addressing such issues, which 
could eventually allow to model the macroscale quantities of interest by a suitable coarse-graining 
procedure.  
We will see that, if one is guided by nonequilibrium 
statistical mechanics and thermodynamics, it is possible to design well-founded multi-scale 
modeling tools that can link microscopic models with macroscopic constitutive equations. 
Such multi-scale modeling tools benefit from recently proposed 
approaches for static coarse graining which are mainly built on potentials of mean force  
\cite{Doros_CGreview,Denton_reviewSoft,Faller_CGreview}. 
For dynamical properties, however, coarse-grained models need to account for dissipative effects 
that arise due to fast degrees of freedom that are eliminated (for example via projection operators)
in favor of the remaining slowly varying ones 
\cite{Briels_CGreview,alexander_workshop,kubobook2,grabert}. 
In a flow situation, these slow dynamical variables depart from their values in the quiescent fluid, while all 
other (faster) degrees of freedom track the evolution of the structural parameters; i.e., they are assumed 
to be in local equilibrium subject to the constraints imposed by the values of the structural parameters 
at all times. 
The proper definition of the set of the state variables, effectively representing the nonequilibrium states 
is the first key to the success of such an approach. Linking the microscopic model with a macroscopic model 
built on these slowly relaxing variables is the second key; as we will discuss in the next sections of this 
chapter, this is best addressed by getting guidance from a nonequilibrium statistical thermodynamics framework 
proposing a fundamental evolution equation for the macroscopic model in terms of the chosen structural 
(dynamical) variables.

%%%%%%%%%%%%%%%%%%%%%%%%%%%%%% review CG %%%%%%%%%%%%%%%%%%%%%%%%%%%%%%%%%%%%%%%%%%%%%%%%%%%%%%
\section{Coarse-grained variables and models} \label{sec:CGmodels}

%\subsection{Mapping to coarse-grained variables} \label{sec:mapping}
We start with a microscopic polymer model, 
whose state is specified by a point in $6N$--dimensional phase space, 
$\z\in\Gamma$ with $\z=(\br_1,\ldots,\br_N;\bp_1,\ldots,\bp_N)$ a short 
notation for the positions and momenta of all $N$ particles. 
The model is described by the microscopic Hamiltonian $H(\z)$ with 
inter- and intra-molecular interactions. 
The coarse-grained model eliminates some of the (huge number of) microscopic 
degrees of freedom. The level of detail that is retained is specified by the 
choice of coarse-grained variables $\x=(x_1,\ldots,x_{\ncg})$ with
\begin{equation} \label{x_def}
 x_k = \ave{\Pi_k} \equiv \int_\Gamma\!{\rm d}\z\, \Pi_k(\z) \rho(\z), \quad k=1,\ldots,\ncg,
\end{equation}
where $\rho(\z)$ denotes the probability distribution on $\Gamma$ and the phase space functions 
$\Pi_k(\z)$ are the instantaneous values of the coarse-grained variables in 
the microstate $\z$. 

Instead of the full, microscopic distribution $\rho(\z)$, the coarse-grained model 
is already specified by the reduced probability distribution 
$p(\x)\equiv\ave{\delta(\x-\bPi(\z))}$. 
Knowledge of $p(\x)$ allows to calculate averages of quantities $a(\bPi(\z))$ via 
$\ave{a(\bPi)}=\int_\Gamma\!{\rm d}\z\, a(\bPi(\z))\rho(\z) = \int\!{\rm d}\x\, a(\x)p(\x)$. 
Instead of $p(\x)$, coarse-grained models are often described by the so-called potentials 
of mean force $\Umf(\x)\equiv -\kb T\ln p(\x)$ which formally replace the Hamiltonian 
in the calculation of equilibrium, canonical averages. 
However, $\Umf$ is an effective free energy difference and therefore depends  
on the thermodynamic state of the system. It contains in general effective many-body interactions 
that arise by partially integrating out microscopic degrees of freedom. 
For recent reviews 
see e.g.~\cite{Denton_reviewSoft,guenza_reviewCG,kleinCGreview,HansenLoewen_effPotreview}.

Different sets of coarse-grained variables have been suggested in the literature and 
are briefly reviewed in the following. 
Usually, one is interested in a drastic reduction of microscopic complexity, so $\ncg\ll 6N$. 
However sometimes, e.g.~for equilibrating atomistic systems, a coarse-grained model with a 
modest reduction ($\ncg$ only a factor 5 to 10 smaller than $6N$) might be useful. 
We emphasize that the mapping (\ref{x_def}) as well as the corresponding probability $p(\x)$ 
is not restricted to equilibrium situations. 
We proceed with a short review of coarse-grained variables and models that capture different 
levels of detail and briefly discuss their static and dynamic properties. 

\subsection{Beads and superatoms} \label{sec:superatoms}
\begin{figure}[htbp]
\centerline{\includegraphics[width=2.01in,height=2.32in]{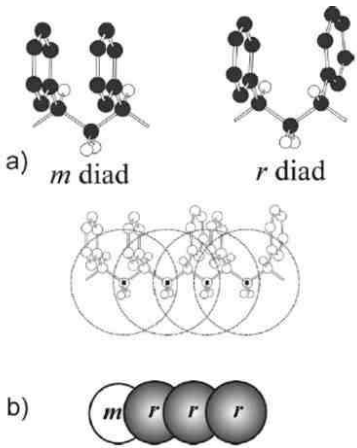}}
\caption{
(a) Polystyrene $m$ and $r$ diads in transplanar conformation (for clarity, hydrogen 
atoms on phenyl rings have been omitted). (b) Illustration of a mapping 
scheme from the atomistic to a coarse-grained structure for PS wherein one 
bead corresponds to an $m$ or $r$ diad. 
[Reproduced with permission from \cite{plathe_reversevinyl}, Figure 1.]
}
\label{sections_3_4_fig1}
\end{figure}

Coarse graining to the level of superatoms 
\cite{Faller_CGpolymer,Faller_CGautomat,Spyriouni,Kamio,plathe_effpotentials,plathe_mapping,plathe_mapvinyl,plathe_reversevinyl,Kremer_CGpolystyrene,Kremer_CGpolystyrene06,KremerAbrams,KremerLeo,plathe_dendrimer,pabloDNA,Kremer_benzLC} 
is a method followed when one wishes to reduce chemical complexity in a polymer chain without loosing 
the chemical identity of the molecule. According to this, a certain number 
of atoms or repeat units along the chain are grouped together into 
``superatoms'' connected by effective bonds and governed by softer or 
smoother effective non-bonded interactions. The resulting (usually linear) 
chain sequence is simpler and amenable to fast thermal equilibration through 
application of state-of-the-art Monte Carlo algorithms slightly modified to 
account for the presence of the few different species along the chain. The 
method has been widely applied to reduce complexity and permit the molecular 
simulation of a number of polymers. Typical examples include polystyrene, 
poly(ethylene terepthalate), polycarbonates, polyphenylene dendrimers, even 
DNA. It involves computing the effective intra and intermolecular potentials 
among superatoms such that the coarse-grained model reproduces as faithfully 
as possible the structural, configurational and thermodynamic properties of 
the original atomistic polymer model. For vinyl chains presenting sequences 
of methylene (CH$_{2}$) and pseudo-asymmetric methyne (-CHR) groups along 
their backbone, the method should also account for the isotactic, 
syndiotactric or atactic stereochemistry of the polymer, based on the 
succession of meso ($m$) and racemo ($r$) diads (see Fig.~\ref{sections_3_4_fig1}).

According to Zwicker and Lovett \cite{Zwicker}, if all interactions with potentials 
$V^{(n)}(\br_1,\br_2,\ldots,\br_N)$, with $\br_i$ denoting 
the position vector of the $i$-th particle, in an $N$-atom molecular system 
consist of $n$-body and lower terms, then the system can be completely 
described by the knowledge of all $n$-order correlation functions 
$g^{(n)}(\br_1,\br_2,\ldots,\br_N)$ and lower. Since, in practice, 
the complete determination of the $n$-point correlation functions is a huge 
task for $N\!>\!2$ and $n\!>\!4$, the calculations are usually limited to correlation 
functions that depend only on a single coordinate. For polymers where 
potential functions are usually separated into intra- and intermolecular 
ones, examples of such correlation functions include typically the radial 
distribution function, the distribution of bond lengths, the distribution of 
bending angles, and the distribution of dihedral angles. The coarse-grained 
potential then should be chosen such that it matches the distributions of 
all possible bond lengths, bond angles, and torsional angles, and of all 
intermolecular pair distribution functions, as extracted from simulations 
with the corresponding atomistic model.

For a distribution function that depends on a single coordinate, the 
corresponding effective potential can be computed through a method called 
iterative Boltzmann inversion method 
\cite{Faller_CGpolymer,Spyriouni,plathe_effpotentials,Soper}, aimed at matching the 
distribution of the relevant degrees of freedom (called target distribution) 
between the chosen coarse-grained model and the initial atomistic model for 
the polymer; the latter are usually extracted from accurate, brute-force MD 
simulations on short homologues. The method uses the differences in the 
potentials of mean force between the distribution function generated from a 
guessed potential and the true distribution function to improve the 
effective potential iteratively. Qualitative arguments for the conditions 
under which convergence should be expected have been discussed by Soper 
\cite{Soper}.

Naive use of the coarse-grained potentials in standard molecular dynamics 
simulations leads to wrong predictions of diffusion and relaxation processes 
\cite{kleinCGreview,plathe_CGwithDPD,plathe_shearPS_CG}. 
A simple, empirical method for relating the dynamics of superatoms at the 
coarse-grained level with the dynamics of true atomistic units at the atomic 
level uses a time rescaling factor \cite{depamaranas,Kremer_DiffusRescaleTime}. 
Within this method, the 
effective potentials are used in the reversible equations of motion of classical 
mechanics to perform standard molecular dynamics simulations and then an effort is 
made to match the mean-square displacements of the relevant structural units in the 
atomistic and coarse-grained models, both in amplitude and slope. 
Noid {\it et al.}~\cite{Noid_CG1,Noid_CG2} formulate 
consistency criteria that should be obeyed when using coarse-grained potentials 
in equilibrium dynamical simulations.

Coarse graining to the level of superatoms has drawn a lot of attention in 
the last years mainly because of the capability to account for the correct 
stereochemical sequence of the repeat units. Despite the success of 
effective pair potentials in reproducing many of the structural properties 
of the corresponding atomistic system, however, their use in actual 
simulations is accompanied by a number of thermodynamic inconsistencies: (a) 
They perform well for the particular physical properties for which they were 
developed. For example, the value of pressure as computed by using the 
virial theorem from the effective potentials optimized with the iterative 
Boltzmann method is higher than what is observed in the atomistic system, 
unless an attractive perturbation potential is added (ramp correction to the 
pressure) and the potential is re-optimized. Given that in integral equation 
methods the mechanical properties of a system (such as pressure, energy and 
compressibility) are fixed by the singlet and pair number densities along 
with proper closures \cite{Zwicker,Henderson_unique,GrayGubbins}, 
such a pressure inconsistency should be 
related to the degree of sensitivity of the site-site pair correlation 
function to the effective pair potential \cite{Soper}. This is in line with the 
simulations of Jain {\it et al.}~\cite{Jain_inverseMC} who showed that, although there is a 
one-to-one correspondence between the structure of a liquid (i.e., the pair 
correlation function) and its pair-wise additive intermolecular potential 
(Henderson's theorem \cite{Henderson_unique}), the convergence of potentials obtained by 
standard inversion procedures is extremely slow: although the repulsive part 
of the potential converges rapidly, its attractive part (to which, e.g., the 
internal energy and pressure are primarily sensitive) converges slowly. 
(b) Effective potentials are in general not transferable; 
they are state-point (e.g., temperature and pressure) dependent. 
In same cases, it has been noticed that temperature changes at about the same 
density do not drastically affect their parameterization \cite{plathe_transferCG,Voth_CGdiffTemp}. 
A newly developed effective force coarse-graining seems to improve transferability 
to other state points \cite{Voth_efforceCG}.
Developing fully self-consistent and transferable 
potentials at any arbitrary level of coarse graining remains still a 
challenge. 
(c) The coarse-grained system is considerably more compressible than the corresponding atomistic one. 
(d) Despite recent efforts, the proper use of coarse-grained potentials for dynamical simulations 
remains unclear. 
In particular the emergence of dissipation due to the coarse-graining step is mostly ignored, or, 
at best, included phenomenologically via some stochastic thermostat as done 
e.g.~in\cite{Guerrault_DPDpoly,plathe_CGwithDPD}. 
For some notable exceptions see \cite{Briels_CGreview}.
Considerable work is definitely 
needed in order to arrive at a thermodynamically consistent description of a 
model system at the two levels of analysis (atomistic and coarse-grained), 
which will eliminate all these undesired symptoms and errors.

\subsection{Uncrossable chains of blobs}

Briels and collaborators 
\cite{BrielsPaddingtransient,BrielsPaddinguncross,Briels_CGchain,Briels_CGdimer} 
proposed a coarse-graining scheme wherein 
chains are sub-divided into a number of subchains of equal length; the 
center-of-mass of each such subchain is taken as the position of a 
corresponding mesoscopic particle called the blob. The blobs are connected 
by springs so that chain connectivity is preserved. Similarly to the 
coarse-graining procedure at the level of superatoms, the method makes use 
of a potential of mean force $\Umf=V({\bf R}^{(n)})$ for the position vectors 
of the $n$ blobs, which ensures that the blob distributions in the atomistic and 
coarse-grained systems are the same. 

In order to describe shear flow effects in a velocity field of the from 
$v_x(\br)=\dot{\gamma}r_y$, Kindt and Briels \cite{BrielsPaddingtransient} proposed making use of 
the SLLOD algorithm \cite{EvansMorriss_SLLOD}. Starting with a Langevin equation, such a method 
results in the following expression for the blob dynamics:
\begin{eqnarray}
\label{eq502}
 M_i \frac{{\rm d}^2{\bf R}_i}{{\rm d}t^2} & = & {\bf F}_i^S 
 -\zeta^{\rm eff}\left( \frac{{\rm d}{\bf R}_i}{{\rm d}t}-\dot{\gamma}P_{iy} \hat{\bf e}_x \right)
 + {\bf F}_i^R \nonumber\\ 
 \zeta^{\rm eff} & = & \zeta +\left[ \sum_i ({\bf F}_i^S \cdot {\bf P}_i - 
 \dot{\gamma}P_{ix} P_{iy}) \right] /\sum_i{\bf P}_i^2  
\end{eqnarray}
where $\hat{\bf e}_x $ denotes the unit vector along the flow ($x$) direction, 
$M_{i}$ is the mass of the $i$-th blob particle, ${\bf R}_{i}$ its position 
vector, ${\bf F}_i^S=-\partial V/\partial {\bf R}_i$ the systematic force on particle $i$, 
$\zeta$ the friction coefficient, and ${\bf F}_i^R $ the random force on particle $i$. 
Since the coarse-grained bonded and nonbonded interactions are so soft that 
unphysical crossing of two bonds would not be prohibited, 
equations (\ref{eq502}) are supplemented with an uncrossability constraint 
of the blob chains. Padding and Briels 
\cite{BrielsKindt,BrielsPaddinguncross} 
realized this constraint by a method which explicitly detects entanglements and  
prevents chain crossings through a geometric procedure. The procedure detects  
possible chain crossings and defines an 
entanglement point \textbf{X} at the prospective crossing site. 
Padding and Briels \cite{BrielsPaddingtransient,BrielsPaddinguncross} 
also proposed some non-trivial order-altering 
moves that lead to creation-removal of entanglements; these are important 
for the best possible realistic treatment of uncrossability constraints in 
simulations with the blob model. 
The Langevin equation of motion (\ref{eq502}) contains the blob friction 
coefficient $\zeta$, whose calculation is not a straightforward issue even under 
equilibrium conditions. Despite this and its simplicity, the blob method has 
been found to capture correctly the viscoelastic properties of polymer melts 
with molecular length several times their entanglement length.
From a numerical point of 
view, the method suffers from large requirements in CPU time, associated 
with the minimization algorithm for the location of entanglements which 
eventually limits simulations to chains made up of a finite number of blobs.

\subsection{Primitive paths}

A method to project atomistically detailed chains to smoother paths was 
proposed by Kr\"{o}ger {\it et al.}~\cite{hco142} through a projection operation that maps 
a set $\{\br_i\}$, $i=1,2,\ldots,N$, of $N$ atomistic coordinates of a 
linear discrete chain to a new set $\{{\bf R}_i\}$ of 
$N$ coarse-grained ones defining a smoother path for the chain that avoids the 
kinks of the original chain but preserves somewhat its topology (the main 
chain contour). The projection involves only a single parameter, $\xi$, whose 
value was obtained by \"{O}ttinger \cite{hco155} by mapping the Porod-Kratky model 
(an atomistic model for a polymer chain) to a smoother chain with a Kuhn 
length equal to the entanglement length. In the limit of infinitely long 
chains, such a mapping suggests that the Kuhn length of the coarse-grained 
chain (the length of a segment between two entanglements) is equal to twice 
the tube diameter. The $\xi$-based method maps a particular chain onto a smoother 
path; however, the reduction of an ensemble of atomistic polymer chains to a 
mesh of primitive paths (PPs) as defined by the Doi-Edwards theory \cite{doibook} 
requires that the projection satisfy not only chain continuity but also 
chain uncrossability. This subtle problem has been addressed only very 
recently through the seminal works of Everaers {\it et al.}~\cite{Everaers_science}, 
Kr\"{o}ger \cite{martin_primitivepath}, and Tzoumanekas-Theodorou \cite{Doros_topos}. 
Nevertheless, the simple $\xi$-based mapping has 
been very helpful in many respects; for example, it has allowed \cite{hco145,Tsolou} to 
successfully calculate the zero shear rate viscosity of model polymer melts 
in the crossover regime from Rouse to entangled. 

The topological analysis of Everaers {\it et al.}~\cite{Everaers_science} 
is based on the idea that PPs can be identified simultaneously for all polymer molecules in a bulk 
system by: keeping chain ends fixed in space, disabling intra-chain 
excluded-volume interactions and retaining the inter-chain ones, and 
minimizing the energy of the system by slowly cooling down toward the zero 
Kelvin temperature. This causes bond springs to reduce their length to zero, 
pulls chains taut, and results in a mesh of PPs consisting of straight 
segments of strongly fluctuating length and more or less sharp turns at the 
entanglement points. The method can be modified 
\cite{Everaers_primitivepath,Kremer_entangled05,Uchida_actin} to preserve 
self-entanglements or to distinguish between local self-knots and 
entanglements between different sections of the same chain.

\begin{figure}[htbp]
\centering
\subfigure[]{\includegraphics[width=2.09in,height=2.17in]{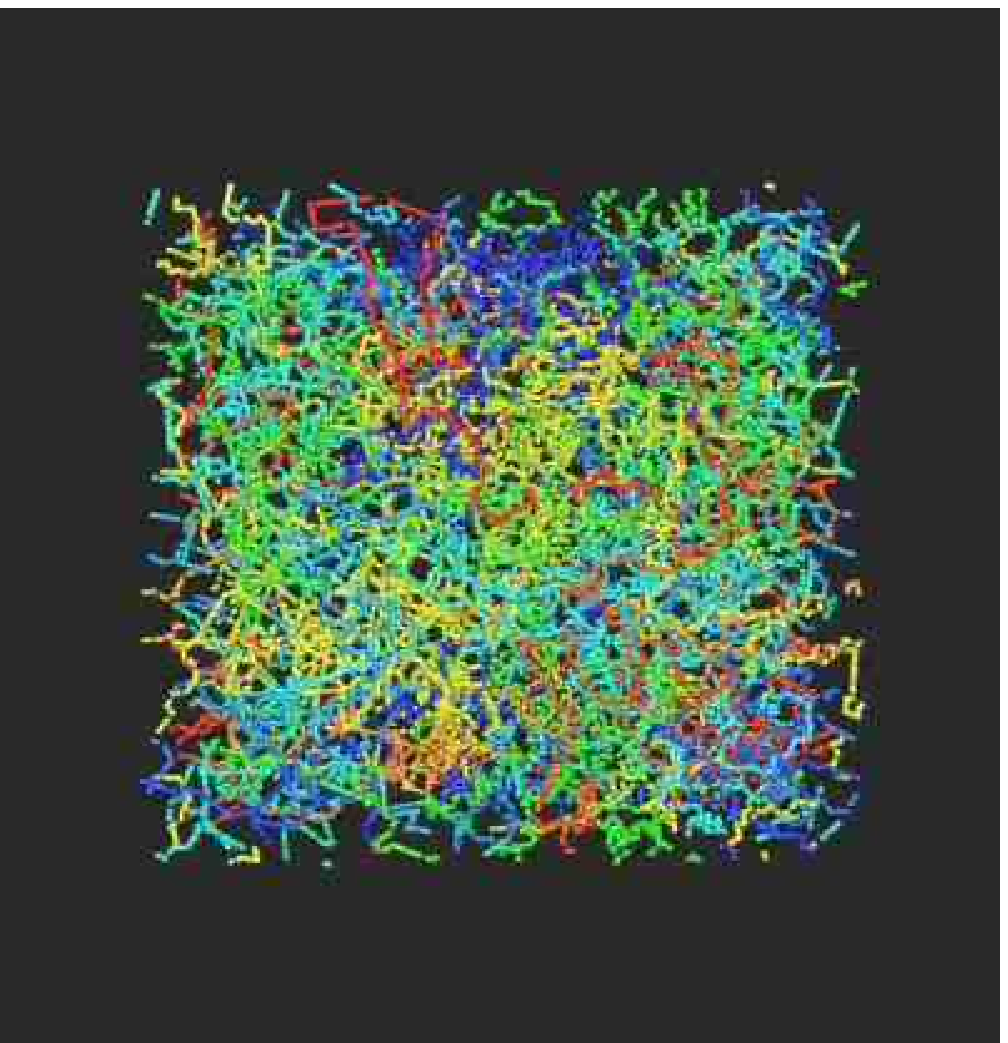}}
\subfigure[]{\includegraphics[width=2.09in,height=2.17in]{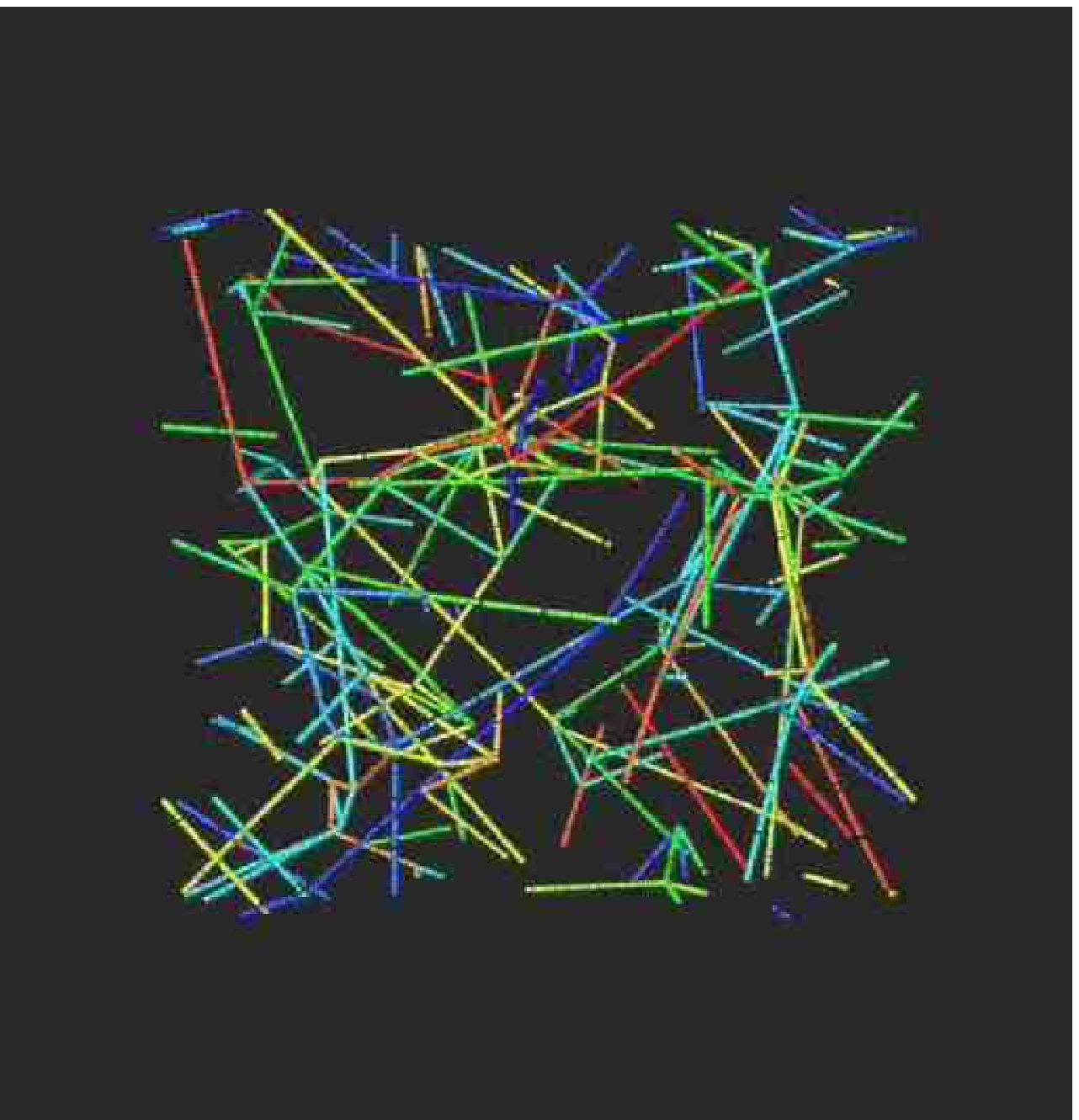}}
\label{sections_3_4_fig2}
\caption{(a) A snapshot of a fully equilibrated atomistic configuration of a 40-chain C$_{500}$ 
trans-1,4-PB melt at 500K and 1atm. (b) The corresponding entanglement 
network mesh as computed with Kr\"{o}ger's method \cite{martin_primitivepath}.}
\end{figure}

Kr\"{o}ger \cite{martin_primitivepath} also presented an algorithm which returns a shortest path 
and the related number of entanglements for a given configuration of a 
polymeric system in 2-D and 3-D space, based on geometric operations 
designed to minimize the contour length of the multiple disconnected path 
(i.e., the contour length summed over all individual PPs) simultaneously for 
all chains in the simulation cell. The number of entanglements is simply 
obtained from the shortest path as either the number of interior kinks, or 
from the average length of a line segment. Application of the algorithm to 
united-atom models of linear polyethylene (PE) \cite{foteinopoulou} allowed the calculation 
of a number of important statistical properties characterizing its PP 
network at equilibrium and helped make the connection with an analytic 
expression for the PP length of entangled polymers by Khaliullin and 
Schieber \cite{jay_pathlength} following earlier works \cite{DoiKuzuu,shanbaglarson}. 
A representative snapshot of 
the entanglement network as computed for a linear trans-1,4-PB polymer 
(40 chains of C$_{500}$ at $T$=450K and $P$=1atm) with Kr\"{o}ger's method is shown in 
Fig.~\ref{sections_3_4_fig2}.

A third methodology for reducing chains to shortest paths has been presented 
by Tzoumanekas and Theodorou \cite{Doros_topos} where topological (chain uncrossability) 
constraints are defined as the nodes of an entanglement network. Through 
their contour reduction topological analysis (CReTA) algorithm, an 
atomistic configuration of a model polymer sample is reduced to a network of 
corresponding PPs defined by a set of rectilinear segments (entanglement 
strands) coming together at nodal points (entanglements) by implementing 
random aligning string moves to polymer chains and hard core interactions. 
In addition to obtaining topological measures for a number of entangled 
polymers, Tzoumanekas and Theodorou \cite{Doros_topos} found that data for the normalized 
distribution of the reduced number of monomers (united-atoms or beads) in an 
entanglement strand are well described by a master curve suggesting a 
universal character for linear polymers. As analyzed by 
Tzoumanekas-Theodorou \cite{Doros_topos}, the master curve is also obtainable in terms of 
a renewal process generating entanglement events stochastically along the 
chain. 

Apart from some algorithmic details, the three methods lead to practically 
similar conclusions as far as the topological state of many entangled linear 
polymer melts (PE, PB, PET and PS) is concerned. A significant result of all 
of them (see, e.g., \cite{Doros_topos}) is that the ensemble average of the number of 
monomers per entanglement strand $\bar {N}_{\rm ES}$ as computed directly from 
the topological analysis is significantly smaller than the corresponding 
quantity $N_{\rm e}$ measured indirectly through 
$N_{\rm e} =N\langle {R^2}\rangle/\langle L \rangle^2$ by assuming that PP 
conformations are random walks. This is due to directional correlations 
between entanglement strands along the same PP, which decay exponentially 
with entanglement strand separation. Therefore PPs are not random walks at 
the length scale of the network mesh size.

\subsection{Other single-chain simulation approaches to polymer melts: slip-link and dual slip-link models}

Doi and Edwards \cite{DoiEdwards78} and Doi and Takimoto \cite{DoiTakimoto} 
have proposed a description 
of an entangled polymer in terms of a slip-link model that can cumulatively 
account for chain confinement and constraint release in a consistent way. 
Slip-links do not represent an entanglement junction in real space; they are 
rather virtual links representing effective constraints whose statistical 
character is determined by other polymers. In the dual slip-link version of 
the model, the slip-link confines a pair of chains (and not a single chain). 
Masubuchi and collaborators 
\cite{marrucci_sliplink,marrucci_sliplink2,Marrucci_tubedilate,Marrucci_comparesliplink} 
generalized the idea by regarding a 
slip-link as an actual link in real space. In their formulation, each 
polymer chain is represented as a linear sequence of entanglement strands 
considered as segments (phantom entropic springs) joining consecutive 
entanglement points (the beads) along the chain. These Rouse-like chains are 
all inter-connected by slip-links at the entanglement points to form a 3-D 
primitive chain network. The system is described by: the number $Z$ of segments 
in each chain, the number $n$ of monomers in each chain segment, and the 
position vectors ${\bf R}_{i}$ of the slip-links or entanglement points. 
These state variables are postulated to obey certain Langevin-type governing 
equations in which the single relevant parameter of the primitive chain 
network is the average value $\langle Z \rangle$ of entanglements per chain. Defining the 
model functions and parameters on the basis of the results obtained from one 
of the three topological analysis discussed above leads to rheological 
predictions that follow quite satisfactorily experimental data for many 
polymer melts in shear but deviations are observed when the model is used to 
describe the elongational rheology of these systems. A generalization of the 
slip-link idea by Schieber and collaborators \cite{jay_sliplink,jay_sliplink2} 
to a full-chain slip-link model with a mean-field implementation of constraint release 
and constant chain friction (as opposed to constant entanglement friction) has 
been shown to provide accurate predictions of the $G'$ and $G''$ spectra 
for many polymer melts (such as PS, PB and PIB).

\subsection{Entire molecules}

In dilute polymer solutions, coarse graining polymer coil or star polymers to a 
a system of interacting soft particles has been explored in 
\cite{Louis_CGpolymsoft,Louis00,Louis01}.
Kindt and Briels \cite{Briels_CGnij,BrielsKindt} proposed such a type of coarse graining 
also for polymer melts where an entire chain is represented as a single particle. 
To account for the presence of entanglements which are considered to be responsible for 
the distinct viscoelastic properties exhibited by polymers, Kindt and Briels 
\cite{Briels_CGnij,BrielsKindt}
introduced a second set of variables, the number $n_{ij}$ of entanglements between 
chains $i$ and $j$. This governs the degree of interpenetration or overlapping of 
two chains whose centers-of-mass are fixed at a given distance. The state of 
the system is thus fully determined by the position vectors ${\bf R}_i$ 
of the centers-of-mass of the $\Nch$ chains and the 
$N_{\rm en}=\Nch(\Nch-1)/2$ entanglement numbers $n_{ij}$. The equilibrium density 
distribution function $\Psi$ for such a system is of the form:
\begin{equation}
\label{eq500}
\Psi \left( {\bf R}^{(\Nch)},n^{(N_{\rm en})} \right)\sim \exp 
\left\{ {-\frac{1}{\kb T}\left[ {\Umf({\bf R}^{(\Nch)})
+{\sum_{i,j} {\frac{1}{2}\alpha \left( {n_{ij} -n_0 
(r_{ij} )} \right)^2} } } \right]} \right\}
\end{equation}
where $\kb$ is Boltzmann's constant, $T$ the temperature, $\Umf$ the potential of mean force, 
the double summation is over all interacting particle pairs, and $\alpha$ is a 
constant determining the strength of the fluctuations around a mean number 
$n_0(r_{ij})$ of entanglements between chains $i$ and $j$. 
$n_0(r_{ij})$ is like an order parameter governing the 
``friction'' felt by each chain and generating restoring elastic forces. 
According to Eq.~(\ref{eq500}), integration over $n_{ij}$ results in a Boltzmann 
distribution for the $\Nch$ coordinates ${\bf R}_{i}$, thus the 
equilibrium statistics of the system is not altered by the introduction of 
the entanglements. Typical expressions for $\Umf$ have been 
discussed by Padding and Briels \cite{BrielsPaddingtransient,BrielsPaddinguncross} 
but also by Pagonabarraga and Frenkel \cite{Frenkel_DPD2000,Frenkel_DPD2001} 
in their derivation of the 
``multi-particle dissipative particle dynamics'' method. 
The method is capable of providing structural 
information only about the radial distribution function $g(r)$ of the 
centers-of-mass of the chains. Representative results for a number of linear 
PE melts revealed a small correlation hole effect at the level of entire 
chains, which is consistent with data reported by 
Mavrantzas-Theodorou \cite{vlasis_freeenergyelong}
through atomistic Monte Carlo simulations. No other signals of local 
structure could be discerned. Clearly, accounting for entanglements (which 
is necessary in order to produce the correct viscoelastic properties) seems 
to have a negligible effect on the structural properties at the level of 
entire chains.

\begin{figure}[htbp]
\centerline{\includegraphics[width=3.24in,height=3.05in]{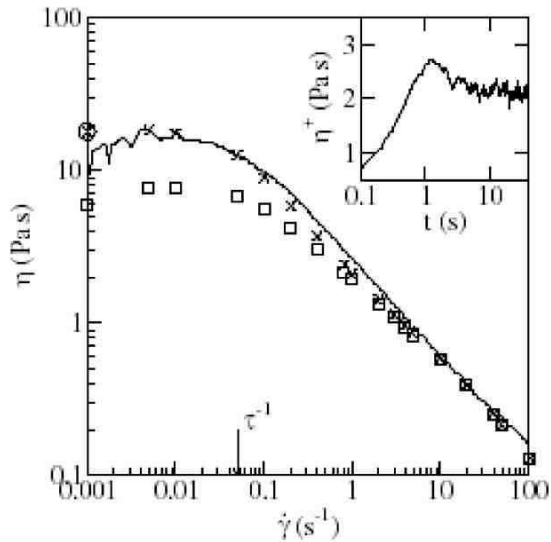}}
\caption{
Viscosity versus shear rate for a typical resin, as obtained from the 
particle model of Kindt and Briels \cite{Briels_CGnij} and 
van der Noort {\it et al.}~\cite{Briels_CGslow}. The 
solid line represents experimental data while the crosses are simulation 
results including all forces in the stress tensor. The squares represent 
viscosities based on the conservative forces only. The circle at the 
vertical axis gives the zero-shear viscosity according to the Green-Kubo 
formula. In the inset, we show the overshoot in the instantaneous value of 
the viscosity at $\dot{\gamma}=$1s$^{-1}$. 
[Reproduced with permission from \cite{Briels_CGslow}, Figure 3.]
}
\label{sections_3_4_fig4}
\end{figure}

The single particle model has been proposed to describe systems where memory 
effects are dominant. This is the case for example of complex fluids 
involving colloidal particles floating in a solvent in which a small amount 
of polymer is also dissolved. In this model, dynamics is described 
\cite{Briels_CGnij,Briels_CGslow} 
by generating (according to standard expressions for Smoluchowski-type 
equations) at every time step ${\rm d}t$ not only a displacement ${\rm d}{\bf R}_i $ in the 
position of each particle $i$ but also a change ${\rm d}n_{ij}$ in the number of 
entanglements. Memory effects are taken into account through transient 
forces: when two particles come together such that temporarily 
$n_{ij}<n_{0}$, their coronas are pushed apart causing a repulsion between 
the two particles. On the other hand, if the coronas are separated enough 
such that temporarily $n_{ij}>n_{0}$, the particles experience attractive 
forces. These phenomena cannot be studied by traditional Brownian Dynamics 
simulations where delta-correlated random displacements are assumed. Figure 
\ref{sections_3_4_fig4} shows the viscosity obtained from such a method referring to a typical 
resin with particles having a hard core diameter of 100 nm. The very same 
model has been successful in describing also shear banding and the chaining 
of dissolved colloids in viscoelastic systems \cite{Briels_CGcoreshell}.

\subsection{Conformation tensor} \label{sec:conftensor}

Based on the idea that, in a flow situation, certain structural variables 
depart from their values in the quiescent fluid while all other (faster) 
degrees of freedom are at equilibrium subject to the constraints imposed by 
the values of the slow variables at all times, coarse-grained models have 
also been developed where chains are described at the level of the 
conformation tensor $\bcvlasis$ \cite{DPL12,mkbook,birdwiest_review,vlasis_elong,Rutledge_QQ}. 
The latter might be defined via the average gyration tensor or via the tensorial product 
of the end-to-end vectors. 
The choice of $\bcvlasis$ among the structural 
parameters marks a description in terms of a \textit{tensorial} variable. 
In addition to a single conformation tensor one can envision a description in terms of many 
(higher-mode) conformation tensors, corresponding to the Rouse or 
bead-spring chain model. In general \cite{vlasis_elong}, from an $N$-mer chain, 
$N(N-1)/2$ different conformation tensors 
$\bcvlasis_{ij}$, $i,j=1,2,\ldots,N-1$ can 
be constructed, each one being identified as a properly dimensionalized 
average dyadic $\left\langle {\bQ_i \bQ_j} \right\rangle $ with $\bQ_i$ 
denoting the connector vector between mers $i+1$ and $i$ along the chain 
\cite{DPL12}.

For systems of entangled polymer chains, 
reptation theory suggests the distribution function of primitive path 
orientations as structural variable, see previous section.
If chains contain long chain branches that can significantly affect the 
rheological response of the system, their contribution to overall system 
dynamics should also be accounted. This is the case, for example with 
H-shaped polymers for which a model built on two structural parameters, the 
tube orientation tensor \textbf{S} from one branch to the other branch along 
the chain, and a scalar quantity $\Lambda$ describing the length of the tube divided 
by the backbone tube length at equilibrium, has been proposed \cite{pompom,hco_pompom}. 
This marks a description at the level of \textit{a tensorial and a scalar}.

The description in terms of a few, well-defined structural parameters (such 
as the conformation tensor, the configurational probability function, and 
the orientation tensor and a scalar) is very appealing because of the 
existence of well-founded models developed under the GENERIC framework of 
non-equilibrium thermodynamics \cite{hcogenbook}. In Section \ref{sec:guidedsimu}, 
we will see that, among other things, this allows building thermodynamically guided multi-scale 
approaches by expanding the equilibrium statistical ensemble to incorporate 
terms involving conjugate variable(s) driving the corresponding structural 
parameters away from equilibrium. The GENERIC formalism here is an extremely 
useful tool, since it can guide us 
in linking the conjugate variable(s) to the applied flow field. 
It is not surprising therefore that this class of coarse-graining procedures 
constitutes one of the most understood and best-founded today.

\subsection{Mesoscopic fluid volumes} \label{sec:mesofluid}

For the numerical simulation of flowing polymers, several mesoscopic models have been 
proposed in the last years that describe polymer (hydro-)dynamics on a mesoscopic 
scale of several micrometers, typically.  
Among these methods, we like to mention Dissipative Particle Dynamics (DPD) \cite{DPD1992}, 
Stochastic Rotation Dynamics (sometimes also called multiparticle collision dynamics) 
\cite{SRD_review}, and Lattice Boltzmann algorithms \cite{saurobook}. 
Hybrid simulation schemes for polymer solutions have been developed recently, combining 
these methods for solvent dynamics with standard particle simulations of polymer beads 
(see e.g.~\cite{SRD_colloidinshear,kaxiras_DNAtranslocation,winkler_MPC04}). 
Extending the mesoscopic fluid models to non-ideal fluids including polymer melts 
is currently in progress 
\cite{Frenkel_DPD2000,Frenkel_DPD2001,Ihle_nonidealMPC,saurobook}.

%\newpage 

%\newpage 

%\newpage 

%%%%%%%%%%%%%%%%%%%%%% GENERIC coarse-graining in general %%%%%%%%%%%%%%%%%%%%%%%%%%%%%%%%%%%%%%%%
\section[Thermodynamic framework of coarse graining]{Systematic and thermodynamically consistent approach to coarse graining: general formulation} \label{sec:CGgeneric}

\subsection{The need for and benefits of consistent coarse-graining schemes} \label{sec:needCG}

Under equilibrium conditions, statistical thermodynamics forms a bridge between thermodynamics 
(whose goal is the understanding and prediction of macroscopic phenomena) and molecular physics 
(which focuses on the intermolecular interactions between the atoms making up the system). 
It provides therefore an interpretation of thermodynamic quantities from a molecular point of view. 
For a number of complex fluids (like colloids, liquid crystals, and polymers), information at a 
mesoscopic level of description (intermediate to molecular and macroscopic ones) is often extremely 
useful in understanding and predicting material behavior 
(see e.g.~Sect.~\ref{sec:CGmodels} and \cite{mkbook}). 
At equilibrium, statistical thermodynamics provides the framework for understanding system properties 
also at these intermediate scales. For example, as mentioned in Sect.~\ref{sec:superatoms},  
effective potentials describing interactions between coarse-grained (pseudo-)particles 
can systematically be derived by integrating out irrelevant degrees of freedom. 

Although extensions to capture dynamics at a mesoscopic level are in progress 
(see Sect.~\ref{sec:CGmodels}), most descriptions in terms of coarse-grained particles 
are so far largely restricted to 
equilibrium situations. For example, the recently proposed time rescaling approach (see 
\cite{Kremer_CGpolystyrene,Kremer_DiffusRescaleTime}) for coarse-grained models does not seem appropriate to 
fully account for the increase in dissipation inherent in any meaningful coarse graining technique. 
We mention the case of hydrodynamic interactions in polymer solution which necessitate a description not in 
terms of a scalar frictional variable but in terms of a tensorial friction matrix. 

Dissipation and friction are more properly accounted for in 
\cite{Briels_CGreview,BrielsPaddingtransient,BrielsPadding,Briels_CGdimer,Briels_CGchain}. 
These authors, however, arrive at a daunting assessment: 
``We therefore conclude that coarse-grained models lack thermodynamic consistency'' \cite{Briels_CGchain}. 
As we will demonstrate below, for appropriately defined coarse-grained models, there is a way to restore 
thermodynamic consistency. So, contrary to the authors of \cite{Briels_CGchain}, we believe that the 
recently introduced GENERIC formalism of nonequilibrium thermodynamics offers a framework for the 
development of true and complete coarse-graining strategies \cite{hco_lessons}, in the sense that: 
(a)  the resulting model is well-behaved and thermodynamically consistent, 
(b) it can be parameterized based on the information provided by a lower resolution model, and 
(c) it can be improved based on microscopic simulations targeted to address the relevant structural 
variables and their dynamic evolution. Of course, a word of caution is in place here: coarse-grained 
models based on the GENERIC framework will rely on a number of strong assumptions 
(inherent to most projection operator based methods) implying a description in terms of a set of 
carefully chosen slowly-evolving state variables. The underlying assumption behind such a description is 
that of the existence of a clear time-scale separation between the evolution of these (slow) variables and 
that of the (eliminated) fast or irrelevant ones. The coarse-grained model of \cite{Briels_CGchain}, 
for example, involves lumping ten beads along a chain into one or two blobs, for which the 
time-scale separation argument is questionable. Their negative conclusion
about the thermodynamic consistency of the model is therefore not 
surprising.

\subsection{Different levels of description and the choice of relevant variables} 
\label{sec:relevantvariables}
Coarse-graining connects (at least) two descriptions of the same system at two different levels of detail: 
a low-resolution level and a higher-resolution level. We focus attention here to the case where the 
high-resolution level is the atomistic one, although this is not necessary \cite{hco_projectors}.

We consider a point in phase space $\z\in\Gamma$, where $\z=(\br_1,\ldots,\br_N;\bp_1,\ldots,\bp_N)$ is a 
short hand notation for the  positions $\br_i$ and momenta $\bp_i$ of all $N$ particles, 
at the microscopic level; this, for example, could be an all-atom or a united atom model or even the 
simpler and computationally more convenient FENE bead-spring model \cite{mkbook}. All these three models are 
classified here as microscopic models due to the absence of dissipation and irreversibility. 
Dynamics at the microscopic model is governed by Hamilton's equation of motion 
\begin{equation} \label{Hamilton}
\dot{\z} =  {\bf J}\cdot \frac{\partial H}{\partial \z}
\end{equation}
where $H(\z)$ is the microscopic Hamiltonian and ${\bf J}$ the symplectic matrix. 
Equivalently, Hamilton dynamics can be formulated by 
$\dot{A}=\{A,H\}$, where $\{A,B\}$ denotes the microscopic Poisson bracket 
between arbitrary functions $A(\z)$ and $B(\z)$. 
We recall that the basic properties of Poisson brackets are their anti-symmetry 
$\{A,B\}=-\{B,A\}$, 
the Leibniz rule 
$\{AB,C\}=A\{B,C\}+\{A,C\}B $, and the Jacobi-identity, 
$\{A,\{B,C\}\}+\{B,\{C,A\}\}+\{C,\{A,B\}\}=0$ \cite{hcogenbook}.

Due to their very long relaxation times, there is a clear gap between the time scales that can be addressed 
in microscopic simulations of polymer melts and the relevant time scales in experimental studies. 
Although this prevents the direct applicability of brute-force microscopic simulations, it renders them  
ideal systems for a comprehensive understanding over multiple time and length scales by embodying the concept of 
multiscale modeling. The first and more important step in this context is the proper choice of the relevant 
variables at the coarser level. For simple fluids, densities of conserved quantities (mass, momentum, and energy) 
are the proper variables to consider if one is interested in hydrodynamic properties. For systems with 
broken symmetries, the corresponding order parameters constitute additional candidates for slow variables 
\cite{ChaikinLubensky}. In the case of complex fluids, however, no general rules are available 
how the appropriate relevant variables should be chosen, and this emphasizes the importance of physical 
intuition \cite{kubobook2,grabert} for the choice of variables beyond equilibrium. 

For polymers, one can be guided by available theoretical models. For example, orientational ordering in 
liquid crystals and liquid crystalline polymers can be described by the alignment tensor within the 
Landau-de Gennes theory \cite{doibook}. 
Birefringence and viscous properties in the case of 
unentangled polymer melts can be addressed by models based on the concept of a conformation 
tensor, see Sect.~\ref{sec:conftensor}. 
For branched polymers, a scalar variable is added to the conformation 
tensor in order to capture additional contributions to the stress tensor \cite{pompom} due to long arm 
relaxations. For entangled polymer melts, reptation theory provides a description in terms of  a 
probability distribution function for the orientation of segments along the primitive path 
\cite{DPL12,doibook}.   

These theories are examples of mesoscopic or macroscopic models that lead to closed-form constitutive 
equations. Furthermore, they can all be described in the context of the single-generator bracket 
\cite{berisbook} or the 
GENERIC \cite{hcogenbook} formalisms of nonequilibrium thermodynamics,   
\begin{equation} \label{generic}
\dot{\x} = {\bf L}\cdot \frac{\delta E}{\delta \x} + {\bf M}\cdot \frac{\delta S}{\delta \x}.
\end{equation} 
In (\ref{generic}), 
$E(\x)$ and $S(\x)$ are the coarse-grained energy and entropy functions, respectively. 
The anti-symmetric operator ${\bf L}$ defines a generalized Poisson bracket 
$\{A,B\}=\frac{\delta A}{\delta \x} \cdot {\bf L} \cdot \frac{\delta B}{\delta \x}$ 
which shares the same properties as the classical Poisson bracket described above. 
The last term in (\ref{generic}) is new compared to Hamiltonian dynamics (\ref{Hamilton}) and describes 
dissipative, irreversible phenomena. 
The friction matrix ${\bf M}$ is 
symmetric\footnote{A more detailed discussion of the Onsager-Casimir symmetry is given in Sect.~3.2.1 of 
\cite{hcogenbook}.}
and positive, semi-definite. 
Together with the degeneracy requirements 
${\bf L}\cdot (\delta S/\delta\x)= {\bf M}\cdot (\delta E/\delta\x)=0$, 
these properties ensure that the 
total energy $E$ is preserved and $S$ is not decreasing in time \cite{hcogenbook}. 

The mesoscopic and macroscopic models come with a number of parameters, 
e.g.~ mean-field potentials, friction coefficients, effective relaxation times, etc., whose connection to 
molecular terms is not straightforward. It is the purpose of thermodynamically-guided, systematic coarse 
graining methods to address this issue. 

\subsection{GENERIC framework of coarse graining}
Coarse graining implies a description in terms of a few, carefully chosen variables after the elimination of 
all irrelevant degrees of freedom. Inevitably, this comes together with entropy generation 
(irreversibility) and dissipation. In the GENERIC formalism \cite{hco_projectors, hcogenbook}, the 
emphasis is then shifted from the fundamental time evolution equation itself to the individual 
building blocks of that theory (\ref{generic}), and opens up the way toward the development of 
consistent coarse-graining strategies. The interested reader is referred here to Ref.~\cite{hco_lessons}. 
More details on the statistical mechanics of coarse graining can be found 
in \cite{kubobook2,hcogenbook,Pep_CGStatMech}; for coarse graining of simple fluids within the 
GENERIC framework, see \cite{hco_simplefluid,henning10}.

\subsubsection{Mapping to relevant variables and reversible dynamics\\} 
\label{sec:mapandreversible}
\noindent
For every microstate $\z$ of the system, the instantaneous values of the relevant variables are defined by a 
set of phase space functions $\bPi(\z)$. The functions $\bPi(\z)$ cannot generally be identified with $\x$; 
they are rather connected with $\x$ through $\x=\ave{\bPi(\z)}$, i.e., as  averages based on
a suitable probability density $\rhox(\z)$ at the microscopic phase space $\Gamma$. Thus, 
the coarse-grained energy $E(\x)$ is obtained from the microscopic Hamiltonian $H(\z)$ by straightforward averaging, 
\begin{equation} \label{ECG_def}
E(\x) = \avex{H(\z)},
\end{equation} 
and, similarly, the coarse-grained Poisson bracket is obtained from the average of the classical Poisson bracket, 
\begin{equation} \label{LCG_def}
\{A,B\}=\frac{\delta A}{\delta x_k} L_{kl} \cdot\frac{\delta B}{\delta x_l}; \quad
L_{kl}(\x) = \avex{\{\Pi_k,\Pi_l\}}.
\end{equation}
Eqs.~(\ref{ECG_def}) and (\ref{LCG_def}) define the reversible part of GENERIC (\ref{generic}) 
in terms of a coarse-grained Poisson bracket 
\cite{dzyalo,ChaikinLubensky}.
The additional terms related to dissipation and increase in entropy have to be accounted for by 
the irreversible contribution to GENERIC (\ref{generic}) and are described just below.

\subsubsection{Irreversibility and dissipation through coarse graining\\}
\noindent
The fact that we do not account explicitly for the irrelevant variables at the level of the GENERIC framework 
leads to entropy increase and additional dissipation at the coarser level of description \cite{hco_lessons}. 

\begin{figure}[h]
\begin{center}
\subfigure[]{\includegraphics[width=6.5cm]{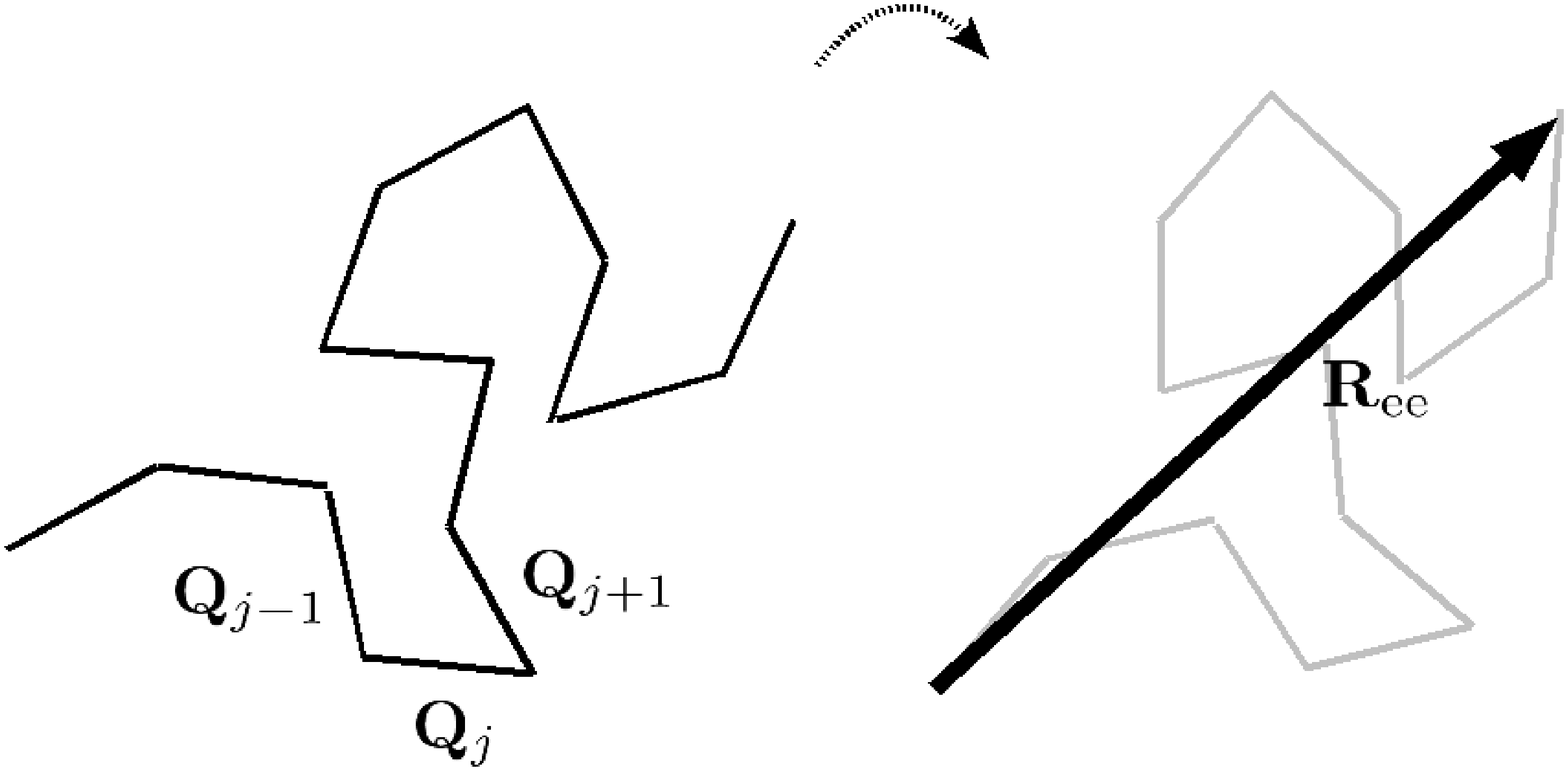}}
\hspace{0.3cm}
\subfigure[]{\includegraphics[width=5.5cm]{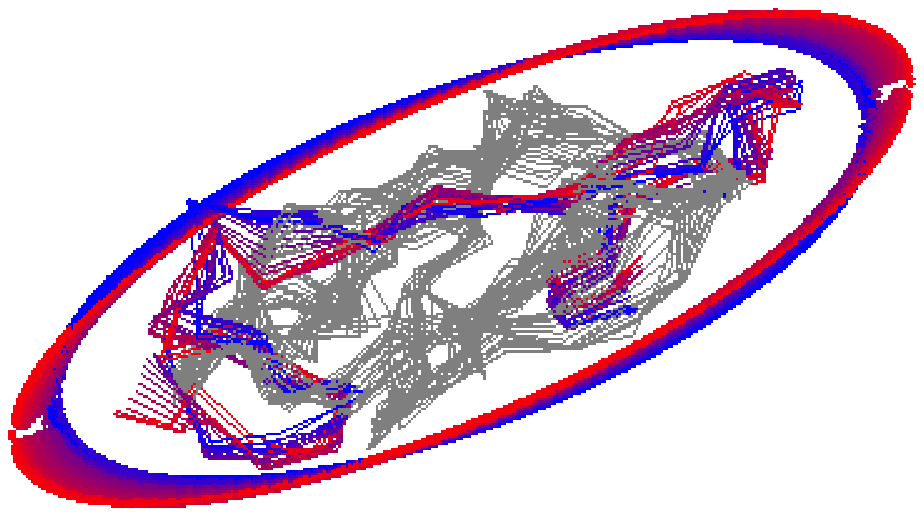}}
\end{center}
\caption{(a) Schematic illustration of freely jointed chain with end-to-end vector $\Ree$. 
(b) Fluctuations of polymer chain in shear flow around stationary state. Ellipses indicate the 
eigenvalues and orientation of eigenvector of $\ave{\Ree\Ree}$. 
[Figure courtesy of M.~Kr{\" o}ger, ETH Z{\" u}rich.]}
\label{randomchain_schematic.fig}
\end{figure}
To illustrate this, let us consider the simple example of a freely jointed chain, shown schematically in 
Fig.~\ref{randomchain_schematic.fig}a. We can describe such a chain by the set of all 
connector vectors ${\bf Q}_j\equiv\br_{j+1}-\br_j$, $j=1,\ldots,N$. 
All admissible configurations with 
$|{\bf Q}_j|=b$  have equal probability. 
If we decide choose the end-to-end vector $\Ree$ as the only relevant variable, then there are in general 
many configurations $\{{\bf Q}_j\}$ which are compatible with $\Ree$. 
The coarse-grained entropy $S(\Ree)$ is a measure of the number of these configurations. One finds that 
the probability of $\Ree$ is Gaussian around $\Ree=0$, which implies that there are many more coiled 
configurations compared to stretched ones.  
The associated entropy $S(\Ree)=S(0)-3\Ree^2/2Nb^2$  
decreases for chains undergoing stretching and leads to a restoring 
force which is known as ``entropic spring'' in coarse-grained polymer models 
\cite{DPL12}. This illustrates the emergence of (additional) entropy 
through coarse graining. 

We turn now to the discussion of the probability distribution $\rho_{\x}(\z)$. In sharp contrast to 
equilibrium statistical mechanics, there are unfortunately no general results for the probability distribution 
of nonequilibrium states. 
Even for nonequilibrium stationary states there are at present only a few results for very 
special model systems available (see e.g.~\cite{Zia_NESSclass}). 
Systems, however, where the time-scale separation assumption holds 
are well described within the quasi-equilibrium approximation that treats the nonequilibrium system as an 
equilibrium one for the present values of relevant variables 
\cite{Jaynes57,zubarevbook,karlinbook}. 
In the generalized microcanonical ensemble, all microstates $\z$ that are compatible with given 
values of the relevant variables $\Pi(\z)$ have equal probability. 
The corresponding entropy is a measure of the number of such microstates $\z$ that are compatible 
with a given coarse-grained state. 
For practical calculations, it is more convenient to pass to the generalized canonical distribution. 
In analogy to equilibrium statistical mechanics, the average values $x_k=\avex{\Pi_k(\z)}$ are 
constrained to prescribed values with the help of Lagrange multipliers $\Lambda_k$. 
The generalized canonical distribution can then be obtained from the maximum entropy principle and reads 
\begin{equation} \label{qea}
\rho_{\x}(\z) = \frac{e^{-\sum_k\Lambda_k \Pi_k(\z)}}{\int_\Gamma\!{\rm d}\z\, e^{-\sum_l\Lambda_l \Pi_l(\z)}}
\end{equation} 
where the $\Lambda_k$'s have to be chosen so as to satisfy 
$x_k=\int\!{\rm d}\z\,\Pi_k(\z)\rho_{\x}(\z)$. 
The quasi-equilibrium entropy associated with Eq.~(\ref{qea}) is 
\begin{equation} \label{qea_entropy}
S(\x) = \kb \sum_k\Lambda_k x_k + \kb\ln \int_\Gamma\!{\rm d}\z\, e^{-\sum_l\Lambda_l \Pi_l(\z)}. 
\end{equation} 
The coarse-grained entropy plays the role of an effective potential for the relevant variables. 
Determining the functional form of $S(\x)$ from (\ref{qea_entropy}) presents a challenge, 
since the explicit expression for the Lagrange multipliers $\Lambda_k(\x)$ is in general unknown.
A successful method for extracting at least partial information on $S(\x)$ has been 
explored in \cite{vlasis_elong} from atomistic simulations of a polymer melt in elongational flow. 
We discuss this issue further in Sects.~\ref{sec:VlasisMC} and \ref{sec:BEMD}.

Having specified the nonequilibrium ensemble and coarse-grained entropy, we finally 
like to discuss the increase of dissipation through coarse graining in more general terms. 
We have seen above that many microstates (values of connector vectors) 
are in general compatible with a given coarse-grained state 
(defined by the value of the end-to-end vector in the above example). 
Conversely, this implies that a coarse-grained state does not uniquely determine the microstate. 
The dynamics on the coarse-grained level has necessarily a stochastic character 
which is known as fluctuations, see Fig.~\ref{randomchain_schematic.fig}b. 
If those fluctuations are correlated in time, 
they have to be accompanied by dissipation, as required by the 
fluctuation-dissipation theorem \cite{kubobook2}. 
These qualitative observations are put into a solid theoretical framework by the 
projection operator formalism \cite{grabert}. 
It should be emphasized that projection operators provide exact relations for any  
set of variables in terms of complicated integro-differential equations. 
Simpler, closed-form equations without a memory integral, however, result only in 
cases when the time-scale of the chosen variables is well-separated from those of the 
irrelevant ones \cite{grabert,Pep_harmonicMarkov}. 
A prominent example where this assumption seems not to be met is the dynamics of glassy polymers, 
where usually mode-coupling approximations for the memory kernel are employed 
\cite{Schweizer_MCTpolymer,Fuchs_MCTpolymer}. 
We here insist on the time-scale separation, 
which severely restricts possible choices of relevant variables where such a separation 
can hold. 
For glassy polymers and glasses, in general, 
an appropriate set of relevant variables is not known at present, 
although some promising first steps have been taken recently 
\cite{hco_glass,ema_nonaffine}. 
These restrictions are the price to pay for a proper coarse-grained description 
with a well-defined entropy and without accounting for memory effects. 
In this case, the dissipation matrix ${\bf M}$ as derived from the projection operator formalism 
reads
\begin{equation} \label{M_int}
M_{kl}=\frac{1}{\kb}\int_0^{\tau_s}\!{\rm d}t\, \ave{\dot{\Pi}_k^{\rm f}(t)\dot{\Pi}_l^{\rm f}(0)}, 
\end{equation}
where $\dot{\Pi}_k^{\rm f}$ is the fast part of the time derivative of the macroscopic variables 
\cite{hco_projectors,zubarevbook,hcogenbook}. 
The separating time scale $\tau_s$ should be chosen large 
enough to comprise all the fast fluctuations that are not captured on the coarse-grained 
level \cite{hcogenbook}. 
Thus, the friction matrix ${\bf M}$ arises due to fast fluctuations that are not resolved at the coarser level. 
The numerical evaluation of the dissipation matrix (\ref{M_int}) for a 
polymer melt is described in Sect.~\ref{sec:BEMD}.

%%%%%%%%%%%%%%%%%%%%%%% Thd guided simulations %%%%%%%%%%%%%%%%%%%%%%%%%%%%%%%%%%%%%%%%%
\section[Thermodynamically guided polymer simulations]{Thermodynamically guided coarse-grained polymer 
simulations beyond equilibrium} 
\label{sec:guidedsimu}

\subsection{GENERIC coarse-graining applied to unentangled melts: foundations} \label{sec:CGconf}
Unentangled polymer melts are usually described in terms of the conformation tensor which provides an 
overall picture of the entire polymer chain. From the point of view of the GENERIC formalism, this implies a 
description where, in addition to the hydrodynamic fields mass $\rho$, momentum $\bmom$, and energy 
density $\en$, the conformation tensor $\bmacro$ is also included in the vector of state variables,
\begin{equation} \label{x_macro}
 \x = (\rho,\bmom,\en,\bmacro).
\end{equation}

Let $\z=(\br_1,\ldots,\br_N;\bp_1,\ldots,\bp_N)$ denote the microstate defined by the positions and momenta
of all particles.
The chosen macroscopic variables $\x=\ave{\bPi(\z)}$ are defined as follows.
The mass density is defined by
\begin{equation} \label{dens_def}
\rho(\br;t) = \ave{\sum_j m_j\delta(\br-\br_j(t))} \equiv \ave{\Pi_\rho},
\end{equation}
where $m_j$ denotes the mass of particle $j$.
Similarly, the momentum density is obtained by
\begin{equation} \label{mom_def}
\bmom(\br;t) = \ave{\sum_j \bp_j\delta(\br-\br_j(t))} \equiv \ave{\bPi_\mom}.
\end{equation}
From $\rho$ and $\bmom$, the macroscopic velocity field $\bv(\br)$ is defined by
$\bv(\br)=\bmom(\br)/\rho(\br)$.
The total energy density can be expressed as
\begin{eqnarray} \label{en_def}
\en(\br;t) & = &
\ave{\sum_j\hat{e}_j\delta(\br-\br_j(t))} \equiv \ave{\Pi_\en},
\end{eqnarray}
where $\hat{e}_j=(1/2)m_j\bu_j^2+\Phi_j$ with
$\bu_j=\bp_j/m_j-{\bf v}(\br_j)$
the peculiar velocity of particle $j$ and
$\Phi_{j}$ the potential energy of particle $j$.
Finally, the additional, internal variable $\bmacro$ is a symmetric, second-rank tensor which is defined by
\begin{equation} \label{macro_def}
\bmacro(\br;t) = \frac{1}{\Nch}\sum_{a=1}^{\Nch}\ave{\hat{\bPi}^{a}\delta(\br-\br_c^{a}(t))}
\equiv\ave{\bPi_{\macro}},
\end{equation}
where $\Nch$ is the number of chains in the system and $N_a=\sum_{j\in I_a}$
the number of particles in chain $a$.
The center of mass of polymer $a$ is denoted by
$\br_c^{a}=N_a^{-1}\sum_{j\in I_a}\br_j$.
The tensor $\hat{\bPi}^{a}$ is a conformation tensor of a single chain and
quantifies the instantaneous, internal structure of polymer $a$.
Examples are the gyration tensor
$\hat{\bPi}^{a}=N_a^{-1}\sum_{j\in I_a}(\br_j-\br_c^{a})(\br_j-\br_c^{a})$ or
the tensor product formed either by the end-to-end vector or the first Rouse mode.

The macroscopic energy $E$ is obtained
by straightforward averaging of the microscopic Hamiltonian, see Eq.~(\ref{ECG_def}).

The resulting expression for the distribution function of the generalized canonical ensemble reads
\begin{equation} \label{qea_womom}
 \rhox(\z) = Z^{-1}\exp{\left[ -\beta p \Vol - \beta \Phi - \blambda\colon\bPi_{\macro}\right]}
\end{equation}
where $\beta=(\kb T)^{-1}$, $\Vol$ the volume occupied by the $N$ particles,
$\Phi$ the total potential energy, $Z$ the normalization integral and $p$ the pressure
(see e.g.~Sect.~8.2.3 in \cite{hcogenbook} and \cite{vlasis_elong,Rutledge_glassyPS} 
where (\ref{qea_womom}) is used in nonequilibrium situations).
The macroscopic entropy associated with the generalized canonical distribution is given by
Eq.~(\ref{qea_entropy}), which here reads
\begin{equation} \label{entropy_lambdac}
 S(\x) = S_0(T,V,N) + \kb\left[ \ln\left(\frac{Z}{V^N}\right) + \beta p V
 + \beta\avex{\Phi} + \blambda\colon\bmacro\right],
\end{equation}
where $V=\avex{\Vol}$ is the average volume
and $S_0$ the entropy of an ideal gas of $N$ particles.
In addition to the usual Lagrange multipliers $\beta$ and $\beta p$ that are
associated with total energy and the volume (for homogeneous density), respectively, 
the additional Lagrange multiplier $\blambda$ is identified as
\begin{equation} \label{lambda_c}
 \blambda = \kb^{-1}\frac{\partial S}{\partial\bmacro}.
\end{equation}
For a numerical calculation of the
Lagrange multiplier $\blambda$ for a model polymer melt see Sect.~\ref{sec:VlasisMC}
and \ref{sec:BEMD}. 

The matrix $L$ defining the coarse-grained Poisson bracket (\ref{LCG_def}) is obtained
by inserting the definitions (\ref{dens_def}-\ref{macro_def}) of the coarse-grained variables
into Eq.~(\ref{LCG_def}).
Details of the straightforward calculations are presented in Ref.~\cite{pi_CGpoly}.

From the degeneracy requirement on the Poisson bracket $\{S,E\}=0$
mentioned in Sect.~\ref{sec:relevantvariables},
one finds that the entropic part of the macroscopic stress tensor has to be of the form
\begin{equation} \label{sigma_macro}
 \bsigma = -p_{\rm eff}{\bf 1} - 2T\bmacro\cdot\frac{\partial s}{\partial\bmacro},
\end{equation}
where $p_{\rm eff}$ is the effective scalar pressure and $s$ the local entropy density.
The same form (\ref{sigma_macro}) has been previously found in \cite{marco,hcogenbook}.

As far as the dissipative bracket and the associated friction matrix ${\bf M}$ are concerned, 
a direct calculation of the fast time evolution $\dot{\Pi}^{\rm f}$ appearing in Eq.~(\ref{M_int})
shows that
$\dot{\Pi}_\rho^{\rm f}=0$,
$\dot{\bPi}_\mom^{\rm f}=\nabla\cdot\hat{\bsigma}^{\rm tot}$,
where $\hat{\bsigma}^{\rm tot}$ is the instantaneous value of the total stress tensor.
The expression for $\dot{\Pi}_\en^{\rm f}$ containing the heat flux and viscous heating
can be found in \cite{pi_CGpoly}.
The integral of the time correlation function of these fast fluctuations
that appears in Eq.~(\ref{M_int}) can in most cases only be determined numerically. 
How these quantities can be extracted from molecular dynamics simulations for a model polymer melt 
is described in Sect.~\ref{sec:BEMD}.

The resulting GENERIC equations (\ref{generic}) for the present choice of relevant variables are
\cite{marco,hcogenbook,pi_CGpoly}
\begin{eqnarray} \label{macrodyn}
\frac{\partial}{\partial t}\rho & = & -\nabla_\beta(v_\beta \rho)\nonumber\\
\frac{\partial}{\partial t}\mom_\alpha & = & -\nabla_\beta(v_\beta\mom_\alpha)
+ \nabla_\beta\sigma_{\beta\alpha}^{\rm tot} \nonumber\\
{\macro}_{\alpha\beta,[1]} & = &
- \frac{1}{T}M_{\macro_{\alpha\beta}\macro_{\mu\nu}}\left[
  \frac{\partial\en}{\partial \macro_{\mu\nu}}
-T\frac{\partial  s}{\partial \macro_{\mu\nu}}\right],
\end{eqnarray}
where ${\bmacro}_{[1]}$ denotes the upper-convected derivative of $\bmacro$, 
${\bmacro}_{[1]}\equiv\partial_t\bmacro + \bv\cdot\bmacro - \bmacro\cdot\bkappa^T - \bkappa\cdot\bmacro$, 
$\bkappa=(\nabla\bv)^T$ the transpose of the velocity gradient, 
and $\en$ and $s$ the energy and entropy density, respectively.
Since we consider in the following only isothermal conditions,
the reader is referred to Refs.~\cite{marco,pi_CGpoly} for the rather
lengthy expression of the internal energy balance.
Furthermore, additional second order dissipative processes appearing in Eq.~(\ref{macrodyn})
are discussed in Ref.~\cite{pi_CGpoly}.

The macroscopic stress tensor appearing in the momentum balance equation (\ref{macrodyn})
is given by
\begin{equation}
 \bsigma^{\rm tot} = -p_{\rm eff}{\bf 1} + 2\bmacro\cdot\left[
  \frac{\partial\en}{\partial \bmacro}
-T\frac{\partial  s}{\partial \bmacro}\right]
- \frac{1}{T}{\bf C}^{(\sigma\sigma)}\colon\bkappa,
\end{equation}
where
${\bf C}^{(\sigma\sigma)}=\int_0^{\taus}\!{\rm d}t\, \avex{\bsigma^{\rm f}(t)\bsigma^{\rm f}(0)}$
is a Green-Kubo formula for the viscosity contribution of fast (on time scale shorter than $\taus$)
stress fluctuations.
This finding is in agreement with
previous simulation studies on bead-spring chain polymer melts
that found it necessary to include a simple fluid background viscosity
in their analysis \cite{Martin_stressopt,Barrat_Rouse}.

%%%%%%%%%%%%%%%%%%%%%%%%%%%%%%%%%%%%%%%%%%%%%%%%%%%
\subsection{Thermodynamically guided atomistic Monte Carlo methodology for generating realistic 
shear flows} \label{sec:VlasisMC}

We discuss here how one, guided by principles of nonequilibrium 
thermodynamics, can make use of the Monte Carlo technique to drive an 
ensemble of system configurations to sample statistically appropriate 
steady-state nonequilibrium phase-space points corresponding to an imposed 
external field 
\cite{vlasis_freeenergyelong,vlasis_elong,Baig_genericMC,Brian_energetic,Brian_energetic2}. 
For simplicity, we limit our discussion to 
the case of an unentangled polymer melt. The starting point is the 
probability density function $\rhox$ of the generalized 
canonical GENERIC ensemble (\ref{qea_womom}) 
for the same set of slow variables (\ref{x_macro}) as in Section \ref{sec:CGconf}. 
Then, following Mavrantzas and Theodorou 
\cite{vlasis_freeenergyelong}, 
we extend the Helmholtz free energy, $A$, of equilibrium systems to 
nonequilibrium systems as
\begin{equation}
\label{eq602}
{\rm d}\left( {\frac{A}{V}} \right)=-\frac{S}{V}{\rm d}T+\mu {\rm d}\left( {\frac{\Nch }{V}} 
\right)-k_B T\blambda\colon{\rm d}\bcvlasis
\end{equation}
where $\mu$ is the chemical potential. 
The last term accommodates the effect of the external field (e.g., a flow) for 
which $\blambda$ represents a nonequilibrium force variable 
conjugate to $\bcvlasis$. 
According to Eqs.~(\ref{qea_womom}) and (\ref{eq602}), 
one can carry out Monte Carlo simulations in the expanded 
$\left\{ \Nch nPT\blambda \right\}$ ensemble exactly as in the corresponding 
$\left\{ \Nch nPT \right\}$ equilibrium ensemble (with $n$ denoting the total number of atoms in 
the system) by assigning non-zero values to the field $\blambda$. 
This is the key point of the new method opening up the way toward sampling 
steady-state nonequilibrium phase points of the system corresponding to a 
given flow field with Monte Carlo by suitably choosing the components of 
$\blambda$. For the case of a simple shear flow, for example, from 
the symmetry property of $\bcvlasis$, we recognize that $\blambda$ 
is to have only four independent non-zero components: 
$\lambda_{xx},\lambda_{xy},\lambda_{yy}$, and $\lambda_{zz}$. 
In order to specify their numerical values for a given 
shear rate $\dot{\gamma}$ (these are needed to be used as input in the 
GENERIC MC simulations), one can resort to the fundamental GENERIC evolution 
law for the set of state variables $\x$. Based on this and Eq.~(\ref{qea_womom}) for 
the definition of $\blambda$, we see that, indeed, we can assign a 
kinematic interpretation to the Lagrange multiplier $\blambda$, 
since for a nonequilibrium system that has reached a steady state, 
Eq.~(\ref{generic}) simplifies to 
\begin{equation}
\label{eq603}
\blambda =-\frac{1}{\kb}{\bf M}^{-1}\cdot {\bf L}(\x)\cdot \frac{\delta E(\x)}{\delta \x}
\end{equation}
For example, for all known conformation tensor viscoelastic models, the 
corresponding evolution equation for the conformation tensor reads:
\begin{equation}
\label{eq604}
\hat{\cvlasis}_{\alpha \beta,[1]} = -\Lambda_{\alpha \beta \gamma 
\varepsilon } \frac{\delta A(\bcvlasis)}{\delta \cvlasis_{\gamma 
\varepsilon } }=-n\kb T\Lambda _{\alpha \beta \gamma \varepsilon } \alpha 
_{\gamma \varepsilon } ;
\quad
\alpha_{\alpha \beta } = \frac{1}{n\kb T}\frac{\delta A(\bcvlasis)}{\delta 
\cvlasis_{\alpha \beta } }.
\end{equation}
where (for simplicity) we have replaced the tensor $\blambda$ with 
the tensor $\balpha$ defined through $\blambda  = -(N/V) \balpha$. 
In Eq.~(\ref{eq604}), $\hat{\cvlasis}_{\alpha \beta}$ 
denotes the upper-convected derivative of $\cvlasis_{\alpha \beta}$ and 
$n$ the chain number density, and the Einstein summation convention has been 
employed for repeated indices. Note also that, in the case considered here, 
the element $M_{44}$ of the ${\bf M}$ matrix in Eq.~(\ref{eq603}) has the form of 
$T\Lambda _{\alpha \beta \gamma \varepsilon }$ 
(see Refs.~\cite{vlasis_elong} and \cite{Baig_genericMC} for details) 
where the fourth-order relaxation matrix $\bLambda$ for 
most single-conformation tensor models can be cast into the following 
general form:
\begin{equation}
\label{eq605}
\Lambda _{\alpha \beta \gamma \varepsilon } (\bcvlasis)=f_1 (I_1 )\,\left( 
 \cvlasis_{\alpha \gamma} \delta_{\beta \varepsilon} + 
 \cvlasis_{\alpha \varepsilon} \delta_{\beta \gamma} + 
 \cvlasis_{\beta \gamma} \delta_{\alpha \varepsilon} + 
 \cvlasis_{\beta \varepsilon} \delta_{\alpha \gamma}  \right) + 
 2f_2 (I_1 )\,\left( 
 \cvlasis_{\alpha \gamma} \cvlasis_{\beta \varepsilon} +
 \cvlasis_{\alpha \varepsilon} \cvlasis_{\beta \gamma} \right)
\end{equation}
where $I_{1}$ is the first invariant of $\bcvlasis$ (i.e., the trace of 
$\bcvlasis$), $\delta$ the unit tensor, and $f_{1}$ and $f_{2}$ 
arbitrary functions of $I_{1}$. With the help of Eqs.~(\ref{eq604}) and (\ref{eq605}), for the 
case of a steady-state flow described by the kinematics
\begin{equation}
\label{eq606}
\nabla \bv =\left( {{\begin{array}{*{20}c}
 0 \hfill & 0 \hfill & 0 \hfill \\
 {\dot{\gamma}} \hfill & 0 \hfill & 0 \hfill \\
 0 \hfill & 0 \hfill & 0 \hfill \\
\end{array} }} \right)
\end{equation}
we find that the form of $\balpha$ that generates shear is
\begin{equation}
\label{eq607}
\balpha =\left( {{\begin{array}{*{20}c}
 {\alpha_{xx} } \hfill & {\alpha_{xy} } \hfill & 0 \hfill \\
 {\alpha_{xy} } \hfill & {\alpha_{yy} } \hfill & 0 \hfill \\
 0 \hfill & 0 \hfill & 0 \hfill \\
\end{array} }} \right)
\end{equation}
Although, however, nonequilibrium thermodynamics has helped us define the 
functional form of $\balpha$, the exact relationship between its three 
non-zero components ($\alpha_{xx}, \alpha_{xy}$, and $\alpha_{yy}$) 
on the applied shear rate $\dot{\gamma}$ 
remains still undetermined. One way to come around this problem is to make 
explicit use of specific expressions for the matrix ${\bf M}$ proposed by 
GENERIC for a viscoelastic model. In such a case, however, the results will 
be model dependent and not representative of the true structure developed in 
the system in response to the applied shear rate $\dot{\gamma}$. Baig and 
Mavrantzas \cite{Baig_genericMC} proposed overcoming this by computing $\balpha$  
iteratively so that, for a given value of $\dot{\gamma}$, the 
resulting average conformation of the simulated melt is the same as that 
predicted through a brute force application of the NEMD method. 

\begin{figure}[htbp]
\centerline{\includegraphics[width=3.00in,height=2.43in]{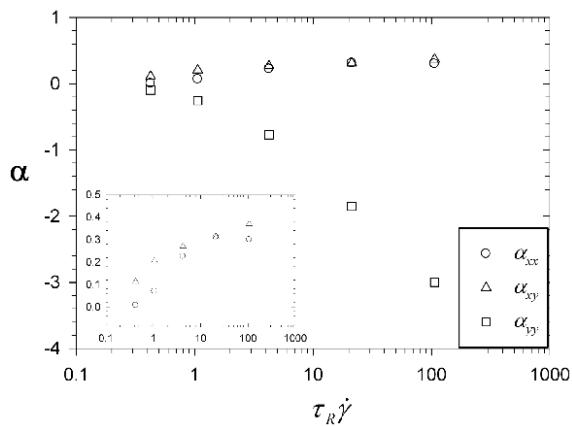}}
\caption{Plot of the thermodynamic force field, $\balpha$, vs.~De number for 
the C$_{50}$H$_{102}$ PE melt ($T$=450K, $P$=1atm). [Reproduced with permission 
from \cite{Baig_genericMC}, Figure 1].}
\label{sec62_fig1}
\end{figure}

\begin{figure}[htbp]
\begin{minipage}[b]{0.5\linewidth}
\centering
\includegraphics[width=2.24in,height=1.75in]{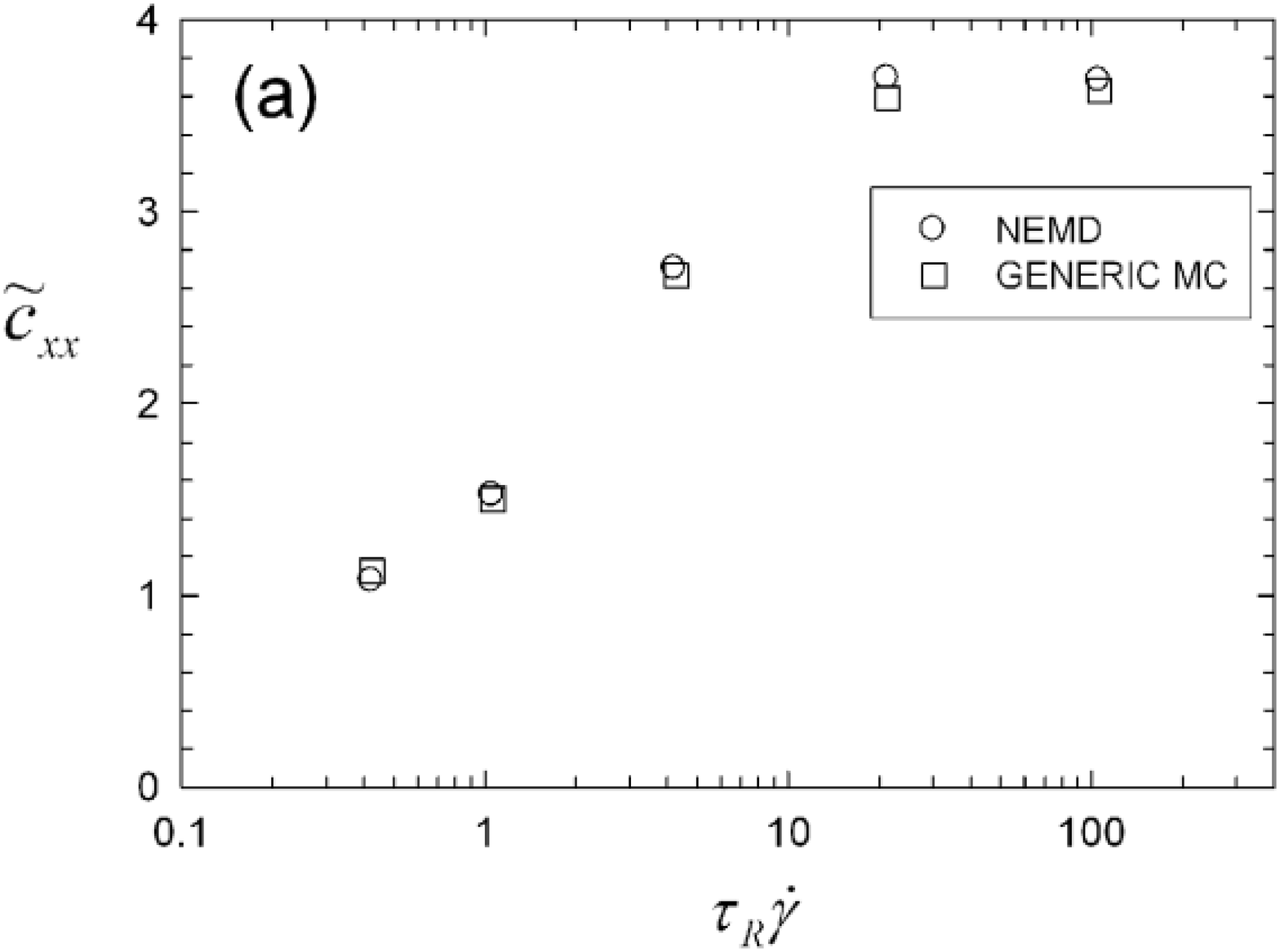}
\end{minipage}
\hspace{0.5cm}
\begin{minipage}[b]{0.5\linewidth}
\centering
\includegraphics[width=2.24in,height=1.75in]{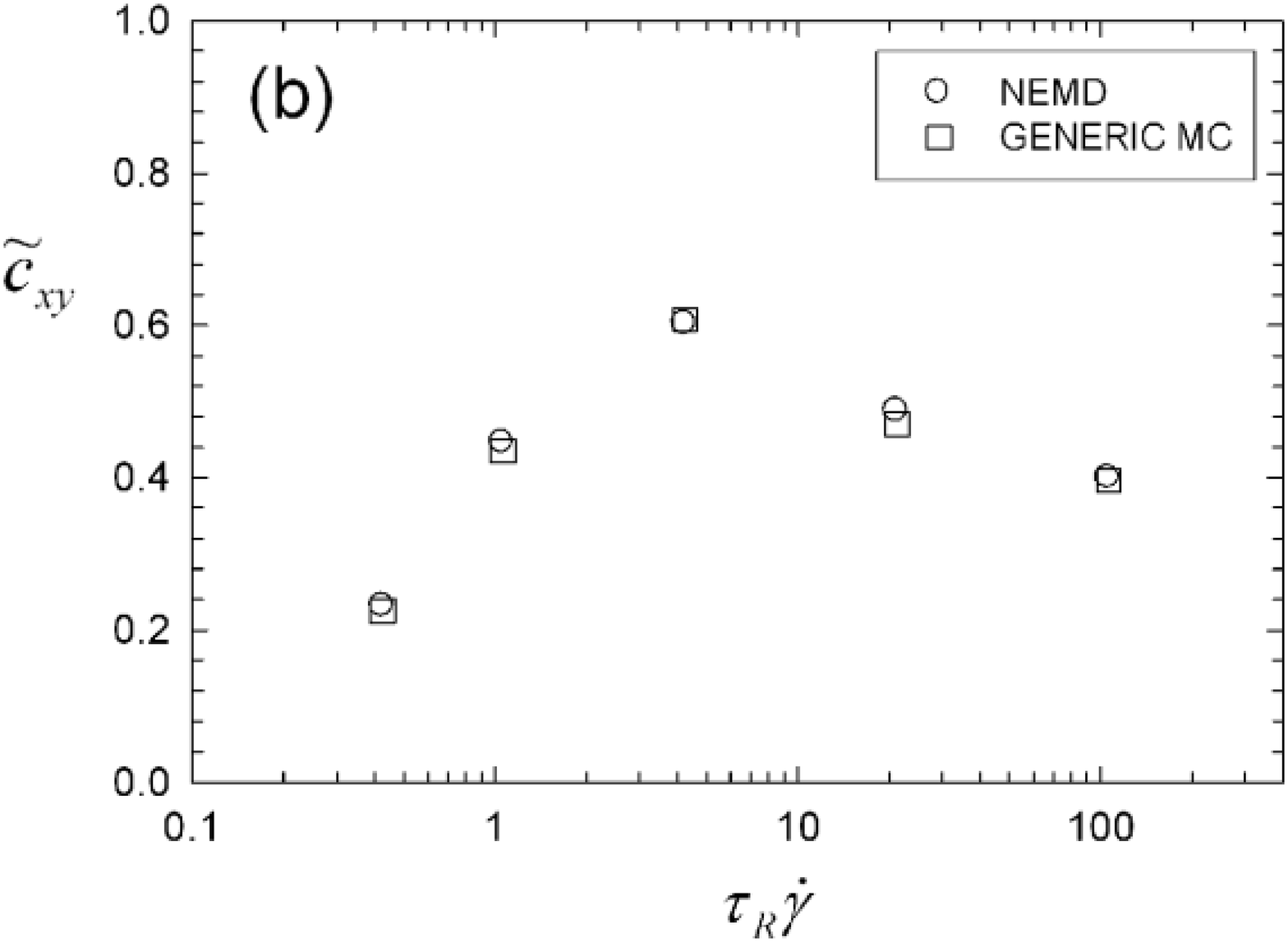}
\end{minipage}
\vspace{0.5cm}
\begin{minipage}[b]{0.5\linewidth}
\centering
\includegraphics[width=2.24in,height=1.75in]{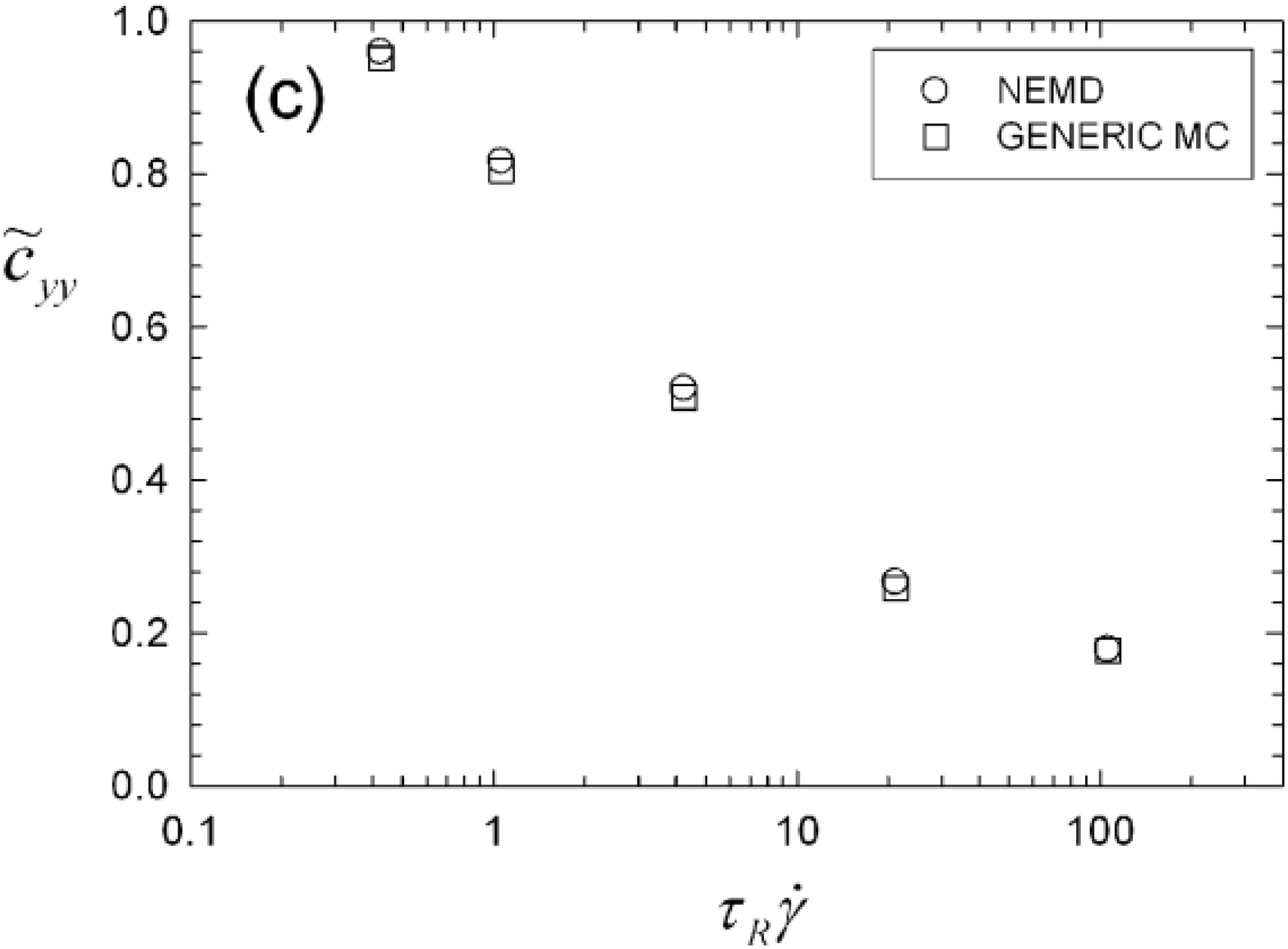}
\end{minipage}
\hspace{0.5cm}
\begin{minipage}[b]{0.5\linewidth}
\centering
\includegraphics[width=2.24in,height=1.75in]{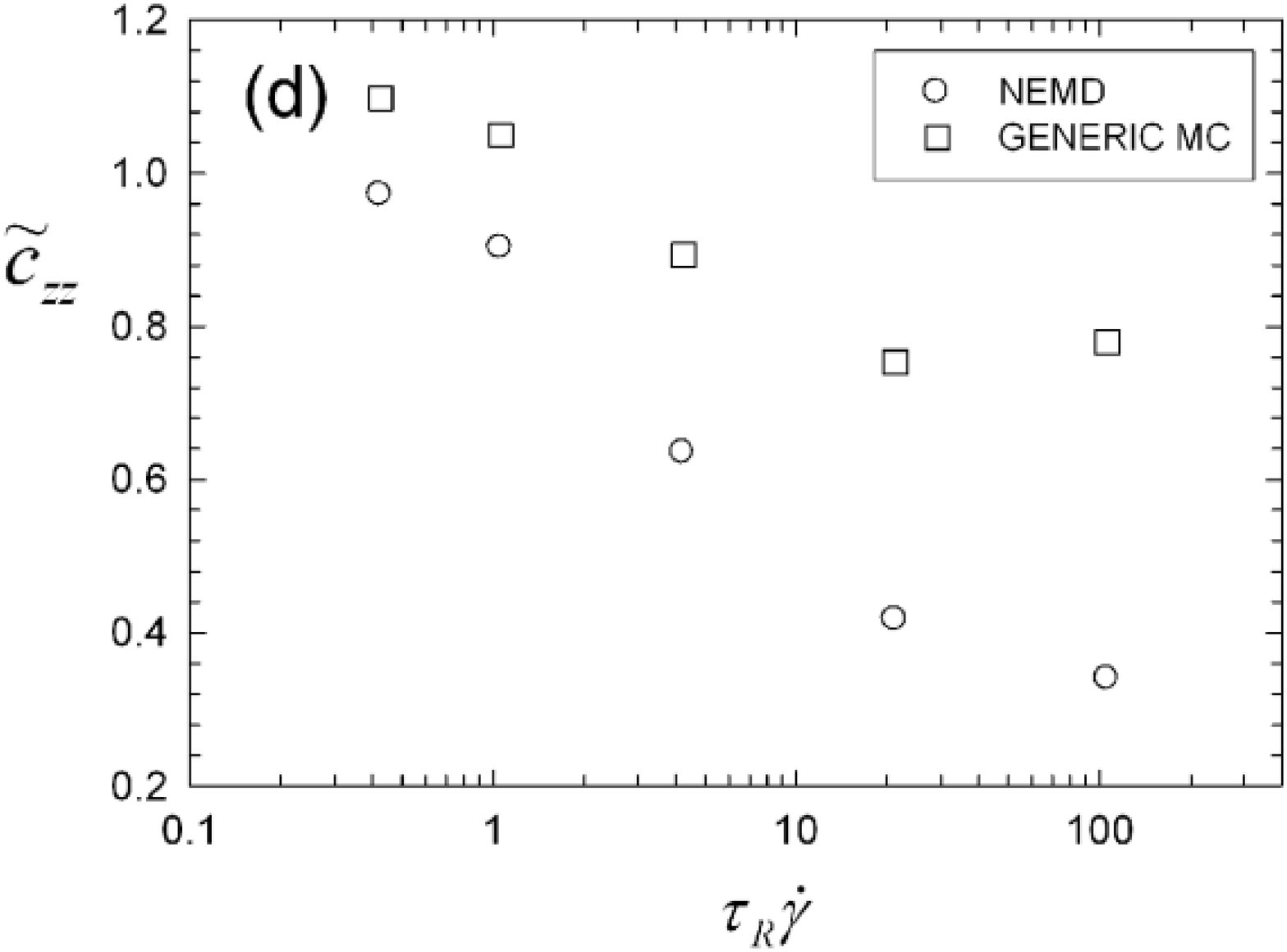}
\end{minipage}
\caption{Comparison of the conformation tensor $\bcvlasis$ components between NEMD 
and GENERIC MC simulations, as a function of De number: (a) 
$\cvlasis_{xx}$, (b) $\cvlasis_{xy} $, (c) $\cvlasis_{yy} $, and (d) $\cvlasis_{zz}$. 
The error bars are smaller than the size of the symbols. [Reproduced with 
permission from \cite{Baig_genericMC}, Figure 1].}
\label{sec62_fig2}
\end{figure}

Baig and Mavrantzas \cite{Baig_genericMC} demonstrated the applicability of such a hybrid 
GENERIC MC - NEMD approach for a relatively short unentangled PE system, 
C$_{50}$H$_{102}$, for different nonequilibrium states corresponding to 
different values of the Deborah number (De). De is defined as the product of 
the imposed shear rate $\dot{\gamma}$ and the longest relaxation (Rouse) 
time of the system, $\tau_R $, at the temperature and pressure of the 
simulation. If $x$ is the flow direction and $y$ and $z$ the velocity gradient and 
neutral directions, respectively, then the three non-zero components of 
$\balpha$ can be computed iteratively so that the values of the 
conformation tensor $\bcvlasis$ for the system  
(at the given value of De) from the GENERIC MC and the NEMD methods 
coincide. Representative results are shown in Figures \ref{sec62_fig1} and 
\ref{sec62_fig2}. Figure \ref{sec62_fig1} 
presents the values of the non-zero components of $\balpha$ that 
were found to reproduce accurately the corresponding nonequilibrium state 
for the simulated C$_{50}$H$_{102}$ system as a function of the imposed De. 
Figure \ref{sec62_fig2}, on the other hand, presents comparisons of the conformation tensor 
between the GENERIC MC simulations (corresponding to the $\balpha$-values 
shown in Figure \ref{sec62_fig1}) and the direct NEMD simulations, confirming 
that $\cvlasis_{xx}, \cvlasis_{xy}$, and $\cvlasis_{yy} $ from the 
GENERIC MC and the NEMD simulations, respectively, superimpose. It is only 
for the $\cvlasis_{zz} $ component of the conformation tensor that Figure 
\ref{sec62_fig2} reveals an inconsistency between the two methods. As argued by 
Baig-Mavrantzas \cite{Baig_genericMC}, this is related to the selection of a zero value for 
the $\alpha_{zz}$ component of $\balpha$, as suggested by the general 
expression, Eq.~(\ref{eq605}). To reproduce exactly also the $zz$-component of 
$\bcvlasis$, a non-zero $\alpha_{zz}$ component should be incorporated in the GENERIC MC 
simulations. This is a significant accomplishment of the new methodology, 
since it suggests that the rather general form of the friction matrix, 
Eq.~(\ref{eq605}), for this conformation tensor family of models is not complete. As demonstrated by 
Baig-Mavrantzas in a recent publication \cite{Baig_PRB}, this can be achieved by including in the 
relaxation matrix $\balpha$ terms beyond the symmetries implied by Eq.~(\ref{eq605}), without violating the 
Onsager-Casimir reciprocity relationships nor the 2nd law of thermodynamics.

The information provided by the GENERIC MC simulations is important in many 
aspects: 

\begin{itemize}
\item The dependence of the components of the tensor $\balpha$ on De is directly related to the 
(nonequilibrium) free energy of system -- see Eq.~(\ref{eq602}). 
Therefore, with the proposed methodology one can accurately calculate the free energy of the 
simulated system by requiring a series of simulations, according to the thermodynamic state points, 
by varying one component of $\balpha$ and fixing the rest and then using thermodynamic 
integration. This can serve as a starting point for developing more accurate viscoelastic models. 
\item The new thermodynamically-guided method can help overcome the problem of long relaxation times 
(and of statistical noise in the Newtonian plateau) faced in brute-force NEMD simulations by providing 
the initial configuration at the relevant nonequilibrium state for a given De.
\item The new method can also be combined with recently proposed coarse-graining simulation strategies 
for long polymer melts to enable the simulation of the viscoelastic properties of high molecular 
weight polymers, comparable to those encountered in practical polymer processing.
\end{itemize}

%%%%%%%%%%%%%%%%%%%%%%%%%%%%%%%%%%%%%%%%%%%%%%%%%%%
\subsection{Systematic time-scale bridging molecular dynamics for flowing polymer melts} \label{sec:BEMD}
We consider again a description of the polymer melt coarse-grained to the level of the conformation tensor. 
The corresponding Poisson bracket is known analytically, see Sect.~\ref{sec:CGconf}.
Same as in Sect.~\ref{sec:VlasisMC}, we investigate the nonequilibrium stationary state of
the polymer melt in a given flow situation,
and therefore face the same problem of solving the stationary GENERIC equation
(\ref{eq603}) self-consistently.
Here, we complete the studies reported in Sect.~\ref{sec:VlasisMC} and consistently determine also the
friction matrix ${\bf M}$ from microscopic fluctuations according to general formula (\ref{M_int}).
Our presentation mainly follows Ref.~\cite{pi_bemd}.

\subsubsection{Systematic time-scale bridging algorithm\\}
The coarse-grained energy and entropy, as well as
the Poisson bracket require only static information and can
therefore be determined very efficiently by Monte Carlo (MC)
simulation methods, see Section \ref{sec:VlasisMC} and 
\cite{Binder_MCguide,Binder_MCMDpoly}. Only the
friction matrix ${\bf M}$ depends on dynamical properties, thus
its numerical evaluation requires dynamical simulation, in
our case Molecular Dynamics (MD). The GENERIC
coarse-graining approach therefore suggests to combine the strengths
of MC and MD simulations in a well-defined way to break the
time-scale gap between microscopic and macroscopic scales.
\begin{figure}
\includegraphics[width=8cm]{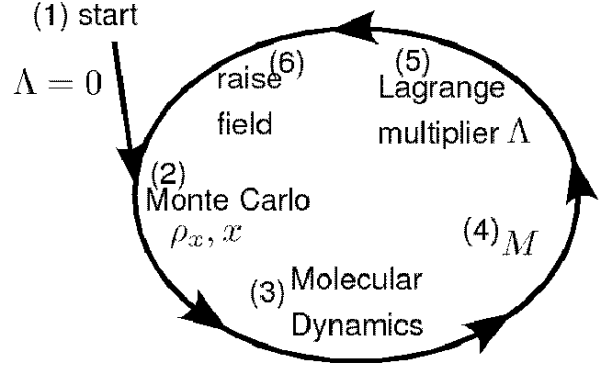} 
\caption{Schematic illustration of systematic time-scale bridging algorithm which consistently
combines Monte Carlo and Molecular Dynamics simulations.}
\label{BEMDscheme.fig}
\end{figure}
In order to implement these ideas consistently we propose a hybrid algorithm \cite{pi_bemd}  
schematically illustrated in Fig.~\ref{BEMDscheme.fig} 
as a general strategy for time-scale bridging simulations based on GENERIC. For the special case 
of flowing, unentangled polymer melts, the algorithm was implemented and tested in Ref.~\cite{pi_bemd}. 
For simplicity and speed of calculations, the classical FENE bead-spring model introduced in 
\cite{loose} was used in these studies, although making use of an atomistic model does not pose 
any extra difficulties.

For this model system subject to a stationary flow with fixed velocity gradient $\bkappa$,
the algorithm illustrated in Fig.~\ref{BEMDscheme.fig} can be implemented as follows.
\begin{enumerate}
\item Start with an equilibrium system with $\bkappa=\blambda={\bf 0}$. 
\item Use a Monte Carlo scheme in order to generate an ensemble of $\nsample$ (typically $\nsample=500$)
independent configurations that are distributed according to the generalized canonical distribution
(\ref{qea_womom}) with the current value of $\blambda$. 
Calculate the value of the relevant variables in this ensemble, $\bx=\avex{\bPi}$.
In order to efficiently generate an ensemble of $\nsample$ such configurations,  
a slight modification of the
Monte Carlo algorithm proposed in \cite{mk_createchains} was used in \cite{pi_bemd}.
The numerical values of the coarse-grained variables $\bx=\avex{\bPi}$ can then be
estimated as the ensemble average $\x=(1/\nsample)\sum_{k=1}^{\nsample} \bPi(\z_k)$ of the 
$\nsample$ configurations $\{\z_k\}$.
In the last step, Maxwellian distributed velocities are assigned to the particles, realizing
equilibrium in momentum space for the present choice of (velocity-independent)
relevant variables.
\item The Monte Carlo generated ensemble is used as initial condition for Molecular Dynamics
simulations of Hamilton's microscopic dynamics. We use a standard velocity-Verlet algorithm
that preserves the symplectic structure to simulate trajectories $\z_k(t)$, $k=1,\ldots,\nsample$,
during a ``short'' time interval $0\leq t\leq \taus$.
The separating time scale $\taus$ is short enough, such that the relevant variables $\x$ do not change
significantly during the MD simulation.
For this reason, the MD part of the simulation does not need any constraints such as thermo- or barostats
nor flow-adapted boundary conditions.
Performing short time, unconstrained, microcanonical molecular dynamics simulations is
one of the great benefits of the present approach as it makes the scheme both highly efficient
and applicable to arbitrary flow situations that -- due to the lack of corresponding
boundary conditions -- could not be simulated so far.
\item From the particle trajectories $\z_k(t)$, we evaluate the friction matrix
${\bf M}$ from Eq.~(\ref{M_int}).
We make use of time-translational invariance to equivalently rewrite
Eq.~(\ref{M_int}) as
\begin{equation} \label{M_delta}
 {\bf M} = \avex{\bbf{\cal M}(z)}, \quad
 \bbf{\cal M}(\z) = \frac{1}{2\kb\taus}\Delta_{\taus}\bPi(\z)\Delta_{\taus}\bPi(\z)
\end{equation}
where
$\Delta_{\taus}\bPi(\z)\equiv\bPi(\z(\taus))-\bPi(\z(0))$
denotes fast fluctuations of $\bPi$ (on the time scale $\taus$).
Equation (\ref{M_delta}) is more convenient for numerical evaluation than (\ref{M_int}).
\item Updated value of the Lagrange multiplier $\blambda$ are calculated from the stationary
GENERIC equation (\ref{eq603}) 
by inverting the symmetric, positive semi-definite matrix ${\bf M}$.
\item The procedure is now repeated until consistent values $\x, {\bf M} , \blambda$ for
given $\bkappa$ are obtained.
Alternatively, one may use an efficient reweighting scheme if $\bkappa$ is changed only slightly and
$\blambda$ is already close to the true value $\blambda\to\blambda+\Deltalambda$.
Then, the explicit form of the generalized canonical distribution can be exploited 
to solve the nonlinear system of equations
\begin{equation}
 {\bf 0} =
   \sum_{k=1}^{\nsample} \left[ {\bf R}_{k} +  \kb \bbf{\cal M}(\z_k)\colon\!\Deltalambda \right]
    w_k, \; w_k \equiv \frac{e^{-\Deltalambda:\bPi(\z_k)} }{\sum_{k'} e^{-\Deltalambda:\bPi(\z_{k'})}}
\label{newMeqabb}
\end{equation}
for $\Deltalambda$.
This first order scheme solves the stationary GENERIC equations (\ref{eq603}), where
${\bf R}_{k} \equiv \bkappa\cdot\bPi(\z_k)+\bPi(\z_k)\cdot\bkappa^T + \kb \bbf{\cal M}(\z_k)\colon\!\blambda$
is the error in the previous value of $\blambda$.
In a shear flow, for example,
Eq.~(\ref{newMeqabb}) represents six equations and six unknowns.
The solution $\Deltalambda$ of (\ref{newMeqabb}) allows one to
calculate the reweighted slow variables and friction matrix,
$\x=\sum_k w_k \bPi(\z_k)$,
${\bf M}=\sum_k w_k \bbf{\cal M}(\z_k)$,
as well as updated Lagrange multipliers,
$\blambda\rightarrow \blambda+\Deltalambda$.
Finally, the flow rate $\bkappa$ is increased, and the procedure is started again, until 
the control parameter space has been swept through. 
\end{enumerate}

With such a scheme, we establish the coarse-grained model along one-dimensional paths in the
parameter space. Choosing, for example, viscometric flows of varying strength $\bkappa$ 
is analogous to the situation encountered in experiments.

\subsubsection{Fluctuations, separating time scale, and friction matrix\\}
\begin{figure}
 \includegraphics[width=8cm]{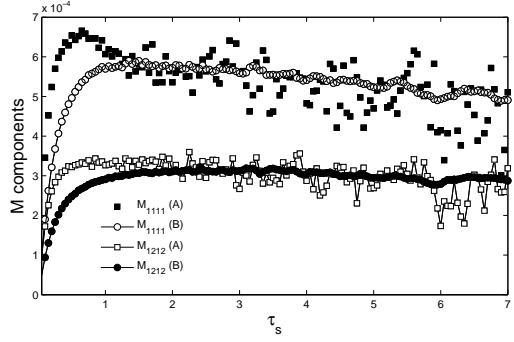}
 \caption{The friction matrix calculated from (A): Eq.~(\ref{M_int}) compared to (B): the values
 obtained from Eq.~(\ref{M_delta}), as a function of the separating time scale $\taus$.
 [Reprinted with permission from \cite{pi_bemd}, Figure 2.]}
 \label{M_taus.fig}
\end{figure}

We have already emphasized several times that ``fast'' but correlated fluctuations give rise to
dissipation on the coarse-grained level of description, which is described here by the
friction matrix ${\bf M}$, Eq.~(\ref{M_int}) or (\ref{M_delta}).
The notion ``fast'' is defined here by times $t$ smaller than the time scale $\taus$,
that separates the evolution of the relevant variables $\bx$ from rapid dynamics of the
remaining degrees of freedom.
The existence of such a time scale (which is equivalent to the crucial assumption of time-scale
separation discussed in Sect.~\ref{sec:CGgeneric}) is not obvious.
Here, we observe that the correlation functions
$C_{kl}(t)=\avex{\dot{\Pi}_k^{\rm f}(t)\dot{\Pi}_l^{\rm f}(0)}$
decay monotonically over a few molecular (Lennard-Jones) time units $\tau$.
This shows that those fast fluctuations are indeed correlated only over short times compared
to typical polymer relaxation times (which are huge relative to $\tau$).
Therefore, we find that the friction matrix, which is proportional to the integral over $C(t)$,
rapidly converges towards a value that is approximately independent of $\taus$ in a broad range
$5\leq \tau/\taus \leq 50$, see Fig.~\ref{M_taus.fig}.

\subsubsection{Results\\}
Before discussing the results obtained with the proposed time-scale bridging algorithm,
we like to mention several consistency checks that can be performed in order to test the
range of applicability of the coarse-grained model.
First, we compare two expressions for the macroscopic stress tensor.
One is the standard virial expression $\bsigma=-V^{-1}\avex{\br{\bf F}}$,
where $\br$ and ${\bf F}$ are the relative position and forces between particles.
The kinetic contribution is found to be negligible in dense systems such as polymer melts
as long as the flow rates are not too high \cite{mkbook}.
Evaluating the expression for the stress tensor $\bsigma$ in the generalized canonical ensemble
leads to the expression
$\bsigma^{\rm p}=-2V^{-1}\kb T \bx\cdot\blambda$ for the (entropic part of the) polymer contribution
to $\bsigma$, see Eq.~(\ref{sigma_macro}).
From Sect.~\ref{sec:CGconf}, we know that $\bsigma$ and $\bsigma^{\rm p}$ differ by a
simple-fluid contribution.
Accounting for this contribution via the non-bonded short range repulsive interactions,
we have verified that the two expressions for the stress tensor agree with each other for
the flow rates studied.
Next, by its definition and the symmetry of $\bPi$, the matrix ${\bf M}$ possesses some basic
symmetries that can be used to test the statistical accuracy of the ensemble averages.
Finally, for the case of simple shear, $\bkappa=\dot{\gamma}{\bf e}_x{\bf e}_y$,
we have used the identity
$(x_{11}-x_{22})x_{12}^{-1}=(\lambda_{11}-\lambda_{22})\lambda_{12}^{-1}$,
that can be derived from the stationary GENERIC equations \cite{Baig_genericMC},
in order to check the consistency of our results.
In our studies, this identity holds within error margins
for the flow rates considered.
We observed that breakdown of this relation at high flow rates signaled problems with the
coarse-grained model as it can no longer capture the relevant dynamical processes
at these elevated rates.

\begin{figure}
 \includegraphics[width=8cm]{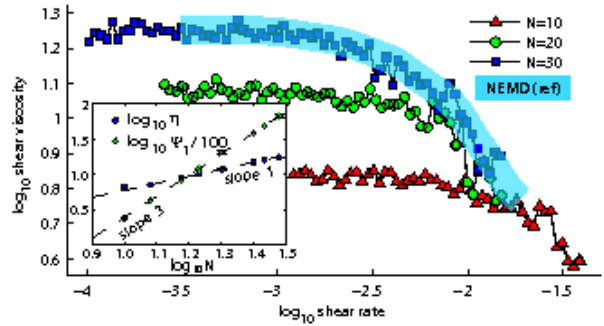}
 \caption{Polymer contribution to the shear viscosity as a function of shear rate for different
 molecular weights. Reference results \cite{loose} obtained with standard NEMD simulations for
 $N=30$ are indicated. 
Inset: Zero shear rate viscosity $\eta_0$ and first normal stress coefficient $\Psi_1$
 as a function of chain length.
 The expected scaling $\eta_0\propto N$ and $\Psi_1\propto N^3$ is observed, as shown
 by the dashed lines.
[Reprinted with permission from \cite{pi_bemd}, Figure 5.]}
 \label{eta_gamma.fig}
\end{figure}

For the case of simple shear flow, we validated the algorithm by reproducing the
chain-length dependence of the zero shear rate viscosity and of the first normal stress coefficient,
which are known in the literature \cite{loose}, see Fig.~\ref{eta_gamma.fig}.
Also the shear rate dependence of the viscosity obtained with the time-scale bridging
algorithm is in very good agreement with standard NEMD results \cite{loose,daivis,daivis_nemd},
as shown in Fig.~\ref{eta_gamma.fig}.
More results can be found in Ref.~\cite{pi_bemd}.
As mentioned above, the flexibility of our time-scale bridging simulations allows us to study arbitrary
flow fields. We therefore could perform the first steady state equibiaxial
simulation for polymer melts. Results for this as well as other elongational flows
can also be found in \cite{pi_bemd}.

%%%%%%%%%%%%%%%%% conclusions and outlook, future challenges %%%%%%%%%%%%%%%%%%%%%%%%%%%%%%
\section[Conclusions and perspectives]{Conclusions and perspectives}

The tremendous multiplicity of length and time scales in polymeric systems clearly calls for 
systematic, multiscale modeling approaches in which a higher-resolution model is consistently 
coupled with a lower-resolution one. In particular, if one is interested in describing relaxation 
processes and structure development under nonequilibrium conditions, most present-day coarse-graining 
strategies based on the use of effective potentials are of limited use since they do not 
account for the additional dissipation and irreversibility accompanying inevitably the elimination 
of fast degrees of freedom in favor of a smaller set of slowly-relaxing structural variables. 
Therefore, {\em thermodynamically guided simulations} are very important and useful, 
where one takes full advantage of the underlying principles of nonequilibrium thermodynamics 
and statistical mechanics. 
There, the emphasis is shifted from the time evolution equations  
(which respect important physical laws such as the Onsager reciprocity relationships for the transport 
coefficients and the 2nd law of thermodynamics) to its four building blocks: the energy $E$, 
the entropy $S$, the Poisson matrix ${\bf L}$ and the friction matrix ${\bf M}$, 
describing the reversible and dissipative contributions to the dynamics.

We have outlined such a methodology for the case of unentangled polymer melts for which, 
guided by network theory approaches to polymer elasticity, the appropriate coarse-grained variable 
$\x$ is the conformation tensor. 
The underlying, microscopic model is simulated by the nonequilibrium molecular dynamics (NEMD) method.
The relevant nonequilibrium state is assumed to be given by a generalized canonical distribution 
incorporating a conjugate variable (the Lagrange multiplier) $\blambda$ to the conformation tensor.
Monte Carlo simulations in this ensemble are employed in order to calculate the values of 
the slow variables $\x$ and the static building blocks $E$ and ${\bf L}$. 
For a given value of imposed flow rates, the Lagrange multiplier can be determined iteratively so that 
the solutions of the micro- and macro-solvers for the coarse-grained structural variables coincide. 
Through this one can compute {\em model-independent} values of the Lagrange multiplier, 
which for a wide range of strain rates (covering both the linear and the nonlinear viscoelastic regimes) 
bring results for the overall polymer conformation from the two models (microscale and macroscale) 
on top of each other. 
We presented two approaches to obtain the missing blocks of the macroscopic model.
In Sect.~\ref{sec:VlasisMC}, the computed values of the Lagrange multiplier are compared 
with those corresponding to specific choices of the friction or relaxation matrix ${\bf M}$ in the 
macroscopic GENERIC model (addressing the chosen structural variable; here the conformation tensor), 
one can identify shortcomings and suggest improvements. And this is the biggest advantage of the 
new framework since the multiscale model proceeds without a priori knowledge of the exact form of 
the macroscopic model. Being built on the GENERIC framework of nonequilibrium thermodynamics, 
what is only needed is just to rely on the nature of the chosen structural variables at the coarse level. 
This, further, emphasizes the significance of the choice of variables in the method.
In the second approach, Sect.~\ref{sec:BEMD}, 
we introduced a novel, low-noise, time-scale bridging strategy for the same system 
(low molecular weight, unentangled polymers) subjected to homogeneous flow fields. 
Through an alternating Monte Carlo-molecular dynamics iteration scheme we were able to obtain 
the model equations for the slow variables. For a chosen flow (including elongational ones), 
the method predicts structural as well as material functions beyond the regime of linear response. 
The method is simple to implement and allows for the calculation of time-dependent behavior through 
quantities readily available from the nonequilibrium steady states. 
In the end, it is only when all three different methodologies (macro-model, micro-model, and the 
macroscopic viscoelastic GENERIC equation bridging them) come together to complement each other that 
the entire multi-scale strategy can be considered as successful. Then, simulation techniques are 
elevated from brute-force computational tools to sophisticated techniques capable of mapping the 
detailed description of the system to a handful of carefully chosen variables whose dynamics 
(time evolution) is also faithfully described by an accurate analytical model.

Future efforts will address other systems such as entangled (linear and branched) polymers where, 
inspired by the corresponding GENERIC formalism, one should resort to a description in terms of 
the orientational distribution function of an entanglement segment along the primitive path of the chain. 

% optional chapter bibliography using BibTeX:
%\bibliographystyle{plain}
\bibliographystyle{unsrt}
%\chapbblname{thdguidedsimupoly}      %% bibliography .bbl file
\bibliography{thdguidedsimupoly}
%\chapbibliography{/home/pilg/TEX/REFS/jabb,refs4wiley,/home/pilg/TEX/REFS/multiscale,/home/pilg/TEX/REFS/generic,/home/pilg/TEX/REFS/polymer} %% database .bib file
\end{document}